\documentclass[a4paper,fleqn]{cas-dc}

\usepackage[square,numbers,compress,sort]{natbib}
\usepackage{csquotes}
\usepackage{xcolor}
\usepackage[colorinlistoftodos,prependcaption]{todonotes}
\usepackage[inline]{enumitem}

\usepackage{xspace}
\usepackage{subcaption}
\usepackage{multicol}
\usepackage[nohyperlinks]{acronym}
\usepackage{float}
\usepackage{amssymb}
\usepackage{pifont}
\newcommand{\cmark}{\ding{51}}%
\newcommand{\xmark}{\text{\ding{55}}}
\usepackage{enumitem}

\def\tsc#1{\csdef{#1}{\textsc{\lowercase{#1}}\xspace}}
\tsc{WGM}
\tsc{QE}
\tsc{EP}
\tsc{PMS}
\tsc{BEC}
\tsc{DE}

\begin{document}
\let\WriteBookmarks\relax
\def\floatpagepagefraction{1}
\def\textpagefraction{.001}

\shorttitle{Automatic Speech Recognition using Advanced Deep Learning  Approaches: A survey}

\shortauthors{H. Kheddar et~al.}

\title [mode = title]{Automatic Speech Recognition using Advanced Deep Learning  Approaches: A survey}                      

\vskip2mm

\author[1]{Hamza Kheddar\corref{cor1}}
[orcid=0000-0002-9532-2453]
\cormark[1]
\ead{kheddar.hamza@univ-medea.dz}
\credit{Conceptualization; Methodology; Data Curation; Resources; Investigation; Visualization;  Writing original draft; Writing, review, and editing}

\author[2]{Mustapha Hemis}
[orcid=0000-0002-6353-0215]
\ead{hemismustapha@yahoo.fr }
\credit{Conceptualization; Methodology; Resources; Investigation; Writing original draft; Writing, review, and editing}

\author[3]{Yassine Himeur}
[orcid=0000-0001-8904-5587]
\ead{yhimeur@ud.ac.ae}
\credit{Conceptualization; Methodology; Resources; Investigation; Writing original draft; Writing, review, and editing}

\address[1]{LSEA Laboratory, Department of Electrical Engineering, University of Medea, 26000, Algeria}

\address[2]{LCPTS Laboratory, University of Sciences and Technology Houari Boumediene (USTHB), P.O. Box 32, El-Alia, Bab-Ezzouar, Algiers 16111, Algeria.}

\address[3]{College of Engineering and Information Technology, University of Dubai, Dubai, UAE}


%
%
%
%
%
%
%
%
%
%
%

\begin{abstract}
Recent advancements in deep learning (DL) have posed a significant challenge for automatic speech recognition (ASR). ASR relies on extensive training datasets, including confidential ones, and demands substantial computational and storage resources. Enabling adaptive systems improves ASR performance in dynamic environments. DL techniques assume training and testing data originate from the same domain, which is not always true. Advanced DL techniques like deep transfer learning (DTL), federated learning (FL), and deep reinforcement learning (DRL) address these issues. DTL allows high-performance models using small yet related datasets, FL enables training on confidential data without dataset possession, and DRL optimizes decision-making in dynamic environments, reducing computation costs.

This survey offers a comprehensive review of DTL, FL, and DRL-based ASR frameworks, aiming to provide insights into the latest developments and aid researchers and professionals in understanding the current challenges. Additionally, Transformers, which are advanced DL techniques heavily used in proposed ASR frameworks, are considered in this survey for their ability to capture extensive dependencies in the input ASR sequence. The paper starts by presenting the background of DTL, FL, DRL, and Transformers and then adopts a well-designed taxonomy to outline the \ac{SOTA} approaches. Subsequently, a critical analysis is conducted to identify the strengths and weaknesses of each framework. Additionally, a comparative study is presented to highlight the existing challenges, paving the way for future research opportunities.

\end{abstract}




\maketitle


\section{Introduction}   \label{sec1}

\subsection{Preliminary}

Advancements in \ac{AI} have significantly improved \ac{HMI}, especially with technologies that convert speech into executable actions. \ac{ASR} emerges as a leading communication technology in \ac{HMI}, extensively utilized by corporations and service providers for facilitating interactions through \ac{AI} platforms like chatbots and digital assistants. Spoken language forms the core of these interactions, emphasizing the necessity for sophisticated speech processing in \ac{AI} systems tailored for \ac{ASR}.
\ac{ASR} technology encompasses the analysis of (i) acoustic, lexical, and syntactic aspects; and (ii) semantic understanding. The \ac{AM} processing includes speech coding \cite{haneche2021compressed}, enhancement \cite{essaid2024advanced}, source separation \cite{luo2021group,essaid2024advanced}, alongside securing speech via steganography \cite{kheddar2019pitch,kheddar2022speech,kheddar2018fourier} and watermarking \cite{yassine2012secure,yamni2022efficient,chen2020specmark}. These components are integral to audio analysis.
On the other hand, the \ac{SM}, often identified as \ac{LM} processing in literature, involves all \ac{NLP} techniques. This \ac{AI} branch aims at teaching computers to understand and interpret human language, serving as the basis for applications like music information retrieval \cite{olivieri2021audio}, sound file organization \cite{wold1996content}, \ac{AT}, and \ac{ED}  \cite{boes2021audiovisual}, as well as converting speech to text and vice versa \cite{tang2021general}, detecting hate speech \cite{plaza2021comparing}, and cyberbullying \cite{mazari2023deep}. Employing \ac{NLP} across various domains enables \ac{AI} models to effectively comprehend and respond to human inputs, unveiling extensive research prospects in diverse sectors.

\begin{table*}
    \centering
{\small \section*{Abbreviations}}
\begin{multicols}{3}
\footnotesize
\begin{acronym}[MAE-AST]
\acro{AAC}{automated audio captioning}
\acro{AC}{actor-critic}
\acro{AI}{artificial intelligence}
\acro{AM}{acoustic model}
\acro{APT}{audio pyramid transformer}
\acro{ASR}{automatic speech recognition}
\acro{AST}{audio spectrogram transformer}
\acro{AT}{audio tagging}
\acro{BERT}{bidirectional encoder representations from transformers}
\acro{CAFT}{client adaptive federated training}
\acro{CER}{character error rate}
\acro{CLDNN}{convolutional long-short term deep neural network}
\acro{CNN}{convolutional neural network}
\acro{CS}{code-switching}
\acro{CTC}{connectionist temporal classification}
\acro{CV}{computer vision}
\acro{DA}{domain adaptation}
\acro{DDPG}{deep deterministic policy gradien}
\acro{DDQN}{double deep Q-network}
\acro{DL}{deep learning}
\acro{DNN}{deep neural network}
\acro{DRL}{deep reinforcement learning}
\acro{DSC}{depthwise separable convolutions}
\acro{DSLM}{domain-specific language modeling}
\acro{DTL}{deep transfer learning}
\acro{ED}{event detection}
\acro{ESPNet}{end-to-end speech processing network}
\acro{FCF}{feature correlation-based fusion}
\acro{FedAvg}{federated averaging}
\acro{FedNST}{federated noisy student training}
\acro{FL}{federated learning}
\acro{FMTL}{federated multi-task learning}
\acro{FR}{form recognition}
\acro{GAN}{generative adversarial network}
\acro{GLDPT}{global-local dual-path Transformer}
\acro{HFL}{horizontal federated learning}
\acro{HMI}{human-machine interaction}
\acro{HMM}{hidden Markov models}
\acro{HSD}{hate speech detection}
\acro{ISTFT}{inverse short-time Fourier transform}
\acro{LLM}{large language model}
\acro{LM}{language model}
\acro{LPC}{linear predictive coding}
\acro{LRF}{low-rank factorization}
\acro{LSTM}{long short term memory}
\acro{MAE-AST}{masked autoencoding audio spectrogram Transformer}
\acro{mAP}{mean average precision}
\acro{MDP}{Markov decision process}
\acro{MFCC}{Mel-frequency cepstral coefficient}
\acro{MHSA}{multi-head self-attention}
\acro{ML}{machine learning}
\acro{MMD}{maximum mean discrepancy}
\acro{MOS-LQO}{mean opinion score-listening quality objective}
\acro{MTL}{multitask learning}
\acro{NLP}{natural language processing}
\acro{non-IID}{non-identically distributed}
\acro{NT}{negative transfer}
\acro{PESQ}{perceptual evaluation of speech quality}
\acro{QCNN}{quantum CNN}
\acro{RER}{relative error rate}
\acro{RNN}{recurrent neural network}
\acro{RTF}{real-time factor}
\acro{S2S}{sequence-to-sequence}
\acro{SARSA}{state–action–reward–state–action}
\acro{SCST}{self-critical sequence training}
\acro{SD}{source domain}
\acro{SE}{speech enhancement}
\acro{SER}{speech emotion recognition}
\acro{SM}{semantic model}
\acro{SNR}{signal-to-noise ratio}
\acro{SOTA}{state-of-the-art}
\acro{SS}{speech security}
\acro{SSAST}{self-supervised audio spectrogram transformer}
\acro{STFT}{short-time Fourier tranform}
\acro{SVD}{singular value decomposition}
\acro{SWBD}{switchboard}
\acro{TBE}{two-branch encoder}
\acro{TD}{target domain}
\acro{TNet}{Transformer network}
\acro{TNR}{true negative rate}
\acro{TPR}{true positive rate}
\acro{TRUNet}{transformer-recurrent-U network}
\acro{TSTNN}{Transformer-based neural network}
\acro{VFL}{vertical federated learning}
\acro{WER}{word error rate}
\acro{WSJ}{wall street journal}
\end{acronym}

\end{multicols}
\end{table*}

Recent advancements in \ac{ASR} have been significantly propelled by the evolution of \ac{DL} methodologies. An extensive range of \ac{DL} models has been developed, demonstrating remarkable improvements and surpassing former \ac{SOTA} achievements \cite{meghraoui2021novel,lin2021speech}. Transformers, a notable innovation within these \ac{DL} approaches, have become a cornerstone in advancing various \ac{NLP} tasks, including \ac{ASR}. Initially conceptualized for \ac{S2S} applications in \ac{NLP}, their success is largely attributed to their adeptness at discerning long-range dependencies and complex patterns within sequential data. A hallmark of Transformer models is their utilization of an attention mechanism, which precisely focuses on specific portions of the input sequence during prediction tasks. This mechanism is particularly effective in \ac{ASR}, facilitating the detailed modeling of contextual nuances and the interconnections among acoustic signals, essential for accurate transcription. Models such as the Transformer Transducer, Conformer, and \ac{ESPNet}, leveraging self-attention and parallel processing, have achieved leading performance in \ac{ASR} tasks. Their robustness across diverse languages further underscores their capability to adapt to a wide range of linguistic features and acoustic variations, making Transformers an exceptionally promising option for enhancing \ac{ASR} systems, surpassing the constraints of conventional models.

The integration of \ac{DL} with its variants in \ac{ASR} introduces substantial challenges, especially concerning its application in natural \ac{HMI}. Despite \ac{DL}'s numerous advantages, it encounters various obstacles. The inherent complexity of \ac{DL} models, which stems from their need for extensive training data to attain high performance, demands significant computational and storage resources \cite{kumar2022novel}. Moreover, the issue of data scarcity in \ac{ASR} reflects the inadequate quantities of training data available for exploiting complex \ac{DL} algorithms effectively \cite{padi2021improved}.
The paucity of annotated data further complicates the development of supervised \ac{DL}-based \ac{ASR} models. Additionally, the presumption that training and testing datasets originate from the same domain, possessing identical feature spaces and distribution characteristics, is often misguided. This mismatch challenges the practical deployment of \ac{DL} models in real-world settings \cite{himeur2022next}.
Thus, the performance of \ac{DL} models may be compromised when faced with limited training datasets or discrepancies in data distribution between training and testing environments \cite{niu2020decade}. These challenges highlight the critical need for adaptive methodologies and improved data management approaches to fully harness the capabilities of \ac{DL} in \ac{ASR} systems.

In an effort to address existing challenges and increase the robustness and flexibility of \ac{ASR} systems, novel \ac{DL} methodologies have been introduced. These include \ac{DTL}, \ac{DRL}, and \ac{FL}, which collectively aim at overcoming difficulties related to the transfer of knowledge, enhancing the generalization capabilities of models, and optimizing training processes. These innovative approaches significantly broaden the operational scope of conventional \ac{DL} frameworks within the \ac{ASR} field.

\begin{figure*}[ht!]
\centering
\includegraphics[scale=0.8]{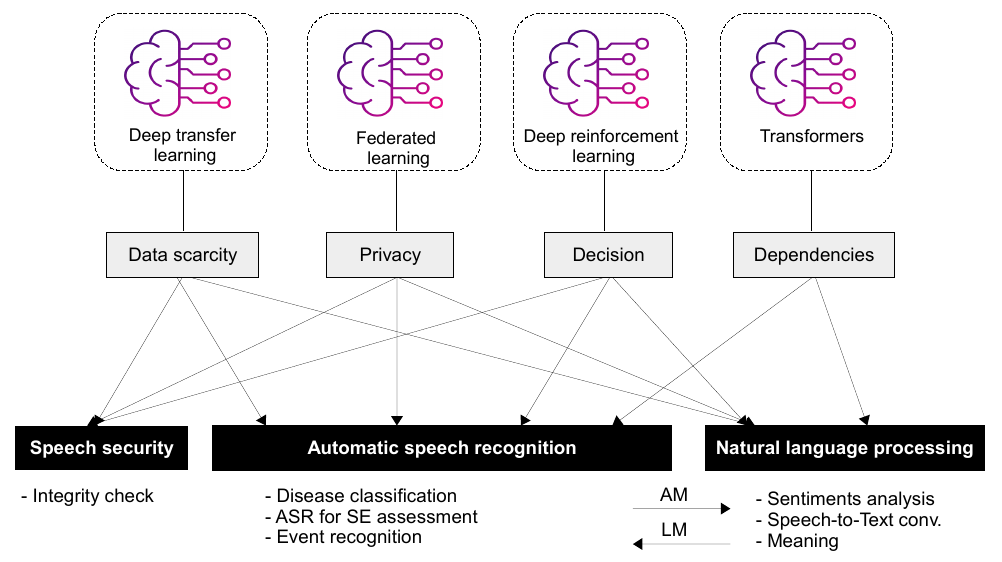}
\caption{ Summary of critical areas in speech processing where \ac{DTL}, DRL, FL, and Transformers can be applied. }
\label{fig:01}
\end{figure*}

Figure \ref{fig:01} highlights critical areas in speech processing where \ac{DTL}, \ac{DRL}, and \ac{FL} can be applied.
Consequently, domains such as \ac{ASR}, \ac{SE}, \ac{HSD}, and \ac{SS} are closely interconnected.
\ac{ASR} provides acoustic parameters to \ac{NLP} for \ac{HSD} task, which in turn provides semantic details to \ac{ASR}. Additionally, \ac{ASR} can be employed in the SS domain as a steganalytic process to verify the integrity of speech \cite{kheddar2019pitch,kheddar2022high}. Furthermore, \ac{ASR} and  \ac{SE} can mutually offer performance feedback.

\subsection{Contribution of the paper}
This article offers an extensive examination of contemporary frameworks within advanced deep learning approaches, spanning the period from 2016 to 2023. These approaches include \ac{DTL}, \ac{DRL}, \ac{FL}, and Transformers, all within the context of \ac{ASR}. To the best of the authors' knowledge, there has been no prior research paper that has intricately explored and critically evaluated contributions in the aforementioned advanced DL-based \ac{ASR} until now.

In recent years, numerous survey papers have been published to assess various aspects of \ac{ASR} models. Some of these surveys concentrate on specific languages, such as Portuguese \cite{de2020survey}, Indian \cite{singh2020asroil}, Turkish \cite{arslan2020detailed}, Arabic \cite{dhouib2022arabic} and tonal languages (including Asian, Indo-European and African) \cite{kaur2021automatic}. Additionally, Abushariah et al.'s review emphasizes bilingual \ac{ASR} \cite{abushariah2022bilingual}. On the non-specific language review front, specific areas within \ac{ASR} have been targeted, including \ac{ASR} using limited vocabulary \cite{fendji2022automatic}, \ac{ASR} for children \cite{bhardwaj2022automatic}, error detection and correction \cite{errattahi2018automatic}, and unsupervised \ac{ASR} \cite{aldarmaki2022unsupervised}. Systematic reviews with a focus on neural networks \cite{dhanjal2023comprehensive} and \acp{DNN} \cite{nassif2019speech} have also been proposed. In another comprehensive review, Malik et al. \cite{malik2021automatic} discussed diverse feature extraction methods, \ac{SOTA} classification models, and some deep learning approaches. Recently, the authors presented an \ac{ASR} review focused on  \ac{DTL} for \ac{ASR} \cite{Kheddar023ASR}. Table \ref{tab:sotareview} presents a summary of the main contributions of the proposed  \ac{DTL} review compared to other existing \ac{DTL} reviews/surveys.

This survey article offers several significant enhancements and additions compared to previous \ac{DTL} surveys. Firstly, it consolidates works that utilize both \ac{DTL} and advanced DL approaches, providing a comprehensive overview of their intersection. Secondly, it provides performance evaluation results of all considered approaches. Thirdly, it includes metrics and dataset reviews used in \ac{DTL} models. Furthermore, it tackles ongoing challenges and consequently proposes future directions. The main contributions of this article can be summarized as follows:

\begin{itemize}
  \item Presenting the background of advanced DL techniques including \ac{DTL}, \ac{DRL}, \ac{FL} and Transformers. 
  Describing the evaluation metrics and datasets employed for validating \ac{ASR} approaches.
  \item Introducing a well-defined taxonomy categorizing \ac{ASR} methodologies based on the domains of \ac{AM} and \ac{LM}.
  \item Identifying challenges and gaps in advanced DL-based \ac{ASR}.
  \item Proposing future directions to enhance the performance of advanced DL-based \ac{ASR} solutions and predicting the potential advancements in the field.
\end{itemize}

\begin{table*}[]
\caption{Contribution comparison of the proposed contribution  against other hand \ac{DTL} review. The tick mark (\cmark) indicates that the specific field has been
addressed, whereas the cross mark (\xmark) means addressing the specific fields has been missed.}
\label{tab:sotareview}
\scriptsize
\begin{tabular}{llm{4cm}m{0.5cm}m{0.5cm}m{0.5cm}m{0.5cm}m{1.5cm}m{1cm}m{1cm}m{1.5cm}m{1.3cm}}
\hline
Refs & Year &  Description of the survey/review & \multicolumn{4}{c}{Advanced DL methods} & Performances  & Metrics &  Dataset & Current  & Future \\ \cline{4-7}
& & & DRL& FL & DTL & Transf.& evaluation & &  review & challe/Gaps & directions  \\\hline
\cite{errattahi2018automatic} & 2018  & ASR review for  errors detection and correction &  \xmark &  \xmark &  \xmark &  \xmark &  \xmark & \cmark &  \xmark & \xmark &\xmark \\
\cite{nassif2019speech} & 2019  & Systematic review on DL-based speech recognition &  \xmark  &  \xmark &  \xmark  & \xmark  &  \xmark & \xmark & \xmark  & \xmark & \xmark \\
\cite{singh2020asroil} & 2020  & ASR survey for Indian languages & \xmark   & \xmark  &  \xmark   & \xmark & \xmark  & \xmark &  \cmark & \xmark & \cmark \\
\cite{de2020survey} & 2020  & ASR survey for Portuguese language & \xmark  &  \xmark  &  \xmark  & \xmark  &\xmark & \xmark & \cmark & \cmark & \cmark  \\
\cite{arslan2020detailed} & 2020  & ASR survey for {Turkish} language & \xmark &  \xmark  &  \xmark & \xmark & \cmark  & \xmark & \cmark  & \xmark & \xmark \\

\cite{malik2021automatic} & 2021  & ASR survey & \xmark  &  \xmark & \xmark & \xmark  &  \xmark & \xmark  & \xmark  &  \cmark & \cmark \\
\cite{kaur2021automatic} & 2021  & ASR survey for tonal languages &   \xmark & \xmark & \xmark  &\xmark  &  \cmark & \xmark & \cmark  & \cmark & \cmark \\
\cite{aldarmaki2022unsupervised} & 2022  & Unsupervised ASR review &   \xmark & \xmark & \xmark  &\xmark  &  \cmark & \xmark &  \xmark & \cmark & \xmark \\
\cite{bhardwaj2022automatic} & 2022  &ASR Systematic review for children &  \xmark & \xmark & \xmark  &\xmark &  \xmark & \xmark & \cmark & \xmark & \xmark  \\
\cite{fendji2022automatic} & 2022  & ASR survey for limited vocabulary & \cmark  &  \xmark  &  \xmark & \xmark  &  \xmark &  \xmark &  \xmark & \xmark & \cmark \\
\cite{dhouib2022arabic} & 2022  & ASR Systematic review for Arabic language & \xmark  &  \xmark & \xmark & \xmark  &  \cmark & \xmark & \xmark  & \cmark & \cmark \\
\cite{abushariah2022bilingual} & 2022  & Bilingual ASR review &  \xmark  & \xmark & \xmark  &\xmark  &  \cmark & \xmark &  \cmark & \cmark & \xmark \\

\cite{dhanjal2023comprehensive} & 2023  & ASR survey on neural network techniques&  \xmark & \xmark &\cmark   & \xmark & \cmark  & \cmark & \cmark & \cmark & \xmark \\

\cite{Kheddar023ASR} & 2023 & ASR based on \ac{DTL} review & \xmark & \xmark & \cmark & \xmark &  \cmark & \cmark  & \cmark & \cmark & \cmark \\

Our & 2024  & ASR review on advanced DL techniques   & \cmark & \cmark  & \cmark  & \cmark & \cmark  & \cmark & \cmark  & \cmark & \cmark \\
\hline
\end{tabular}
\end{table*}

\subsection{Review methodology} \label{sec1.3}

The methodology for the review is delineated in this segment, encompassing the search strategy and study selection. Inclusion criteria, comprising keyword alignment, creativity and impact, and uniqueness, are explicated, collectively influencing the formulation of the paper's quality assessment protocol. To locate and compile extant advanced DL-based \ac{ASR} studies, a thorough search was executed on renowned publication databases recognized for hosting top-tier scientific research articles. The exploration encompassed Scopus and Web of Science. Keywords were extracted and organized from the initial set of references through manual analysis. Employing "theme clustering," these publications were sorted based on keywords found in the "Abstract," "Title," and "Authors keywords." The outcome of this process yielded the formulation of the following query:

\begin{center}
References=FROM "Abstract" || "Title"|| "Authors keywords" SELECT\big( Papers WHERE keywords= (\ac{ASR} || \ac{NLP}) \& ( \ac{DTL} || \ac{DRL} || \ac{FL} || Transformers)\big).
\end{center}

\noindent The symbols || and \& denote OR and AND logical operations, respectively. The evaluation of publications considered the innovation level in \ac{ASR}, the study's quality, and the contributions and findings presented. This review exclusively encompassed research contributions that were published within the timeframe of 2016 to 2023.

\subsection{Structure of the paper}

\begin{figure*}
    \centering
    \includegraphics[scale=0.55]{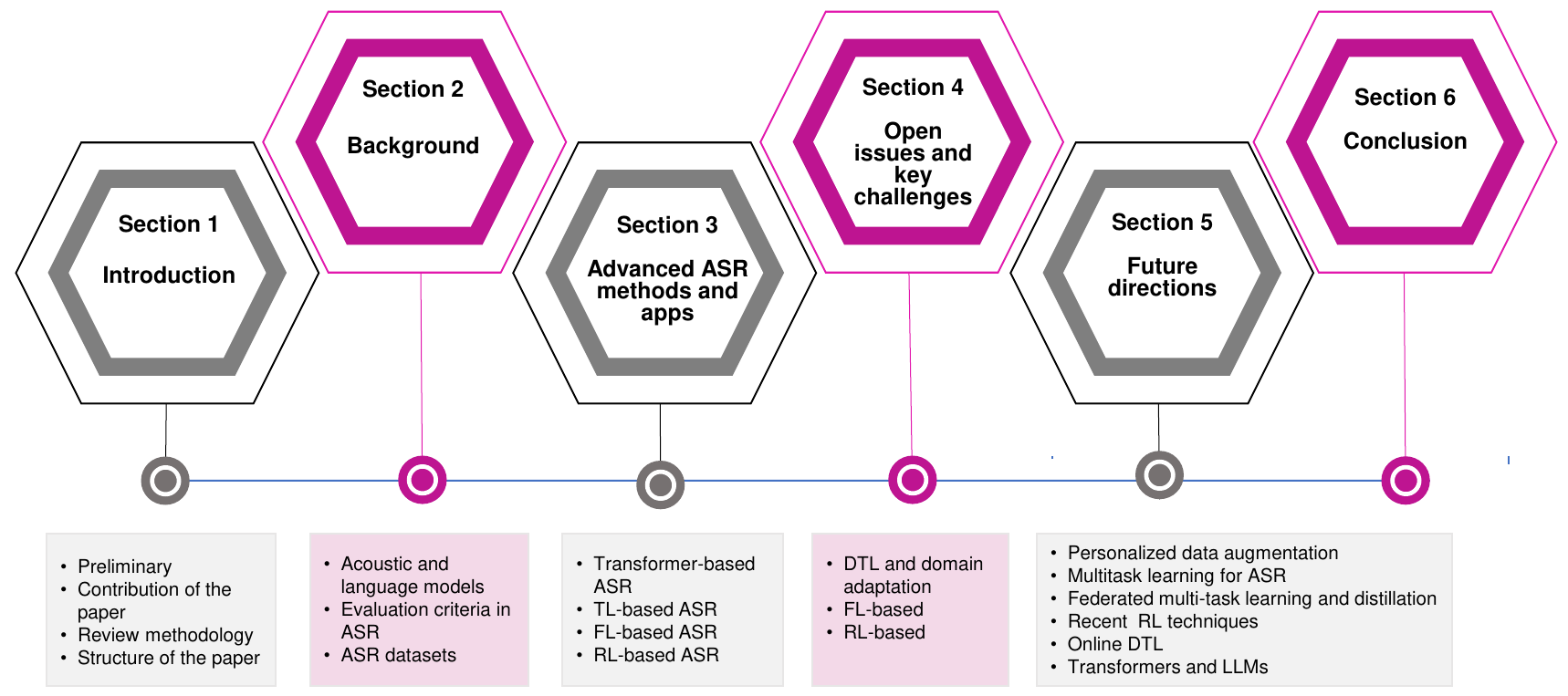}
    \caption{Survey roadmap: A guide for navigating paper sections and subsections.}
    \label{fig:roadMap}
\end{figure*}

This paper is structured into six sections. The current section provides an introduction to the paper. {Section \ref{sec2}} providing background on \ac{AM} and \ac{LM}, and reviewing evaluation metrics and datasets utilized in \ac{ASR}. Moving forward, Section \ref{sec3} delves into a comprehensive review of recent advancements in \ac{ASR} utilizing advanced \ac{DL} approaches, including Transformers, \ac{DTL}, \ac{FL} and \ac{DRL}. Sections \ref{sec4} and \ref{sec5} respectively address the existing challenges and future directions concerning advanced DL-based \ac{ASR}. Finally, Section \ref{sec6} presents concluding remarks summarizing the key findings of the paper. {Figure \ref{fig:roadMap} presents a structured roadmap, offering a comprehensive guide to assist readers in navigating through the various sections and subsections of the paper.}

\section{Background}
\label{sec2}
\subsection{Acoustic and language models} 
\begin{figure}[ht!]
\centering
\includegraphics[scale=0.7]{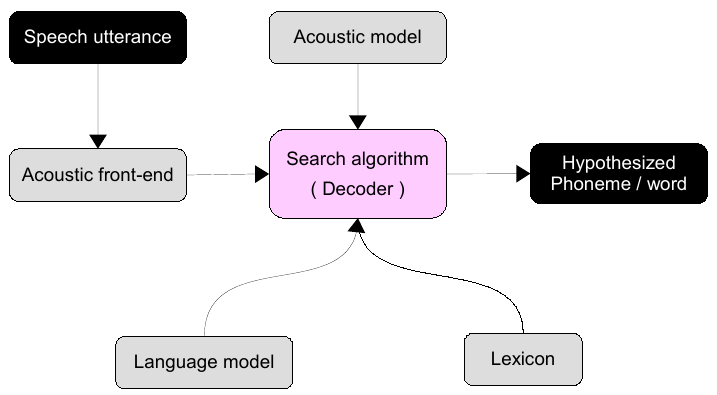}
\caption{Diagram illustrating the end-to-end framework for \ac{ASR}.}
\label{fig:ASR}
\end{figure}

The \ac{AM} is in charge of capturing the sound characteristics of different phonetic units. This involves generating statistical measures for characteristic vector sequences from the audio waveform. Various techniques, such as \ac{LPC}, Cepstral analysis, filter-bank analysis, \acp{MFCC}, wavelet analysis, and others, can be used to extract these features \cite{filippidou2020alpha}. In the processing stage, a decoder (search algorithm) uses the acoustic lexicon and \ac{LM} to create the hypothesized word or phoneme. You can see the overall process illustrated in Figure \ref{fig:ASR}.

\Acp{LM} provide probabilities of sequences of words, crucial for \ac{ASR} systems to predict the likelihood of subsequent words in a sentence \cite{suzuki2023constructing}. A domain-specific LM is trained on text data from the target domain to capture its unique vocabulary and grammatical structures \cite{yang2023generative}. For n-gram models, this involves calculating the conditional probability of a word given the previous $n-1$ words \cite{dong2023speech}:
\begin{equation}
\scriptsize{
    P(w_n | w_{n-1}, w_{n-2}, \ldots, w_{n-(n-1)}) = \frac{C(w_{n-(n-1)}, \ldots, w_n)}{C(w_{n-(n-1)}, \ldots, w_{n-1})}}
\end{equation}

In the context of \ac{ASR}, the \ac{LM} complements the \ac{AM} by providing linguistic context. The combined probability from the \ac{AM} and the \ac{LM} helps in determining the most likely transcription for a given audio input during the decoding process. The frequently utilized \ac{LM} in \ac{ASR} systems is the backoff n-gram model.

\subsection{Evaluation criteria in ASR}

To assess the effectiveness and suitability of \ac{ASR} techniques, researchers have employed diverse methods. Some of these encompass well-established \ac{DL} metrics, including accuracy, F1-score, recall (sensitivity or \ac{TPR}), precision (also known as positive predictive value), and specificity (commonly referred to as \ac{TNR}) \cite{kheddar2023deepIDS}. These metrics serve as crucial evaluation criteria for experimental outcomes, as evidenced in studies such as \cite{li2021insight, karaman2021robust, ramadan2021detecting}. Additionally, there are \ac{ASR}-specific metrics, which are detailed in Table \ref{tab:metrics}.

\begin{table*}[]
\scriptsize
\begin{tabular}[!t]{m{15mm}m{50mm}m{100mm}}
\caption{An overview of the metrics employed for evaluating \ac{ASR} methods.}
\label{tab:metrics} \\
\hline
Metric & Formula & Description   \\ 
\hline
 
\acs{WER} & \(\displaystyle
    \mathrm{\frac{S+D+I}{N}=\frac{S+D+I}{H+S+D}}. \) &   The \ac{WER} serves as a frequently utilized metric to assess the performance of Automatic Speech Recognition (\ac{ASR}). It is computed by determining the ratio of incorrectly recognized words to the overall number of processed words \cite{yu2021enhancing,lin2021speech,bai2021fast}. In the given context, $\mathrm{I, D, S, H}$, and $\mathrm{N}$ denote the quantities of insertions, deletions, substitutions, hits, and input words, respectively. Instead of \ac{WER}, the \ac{CER} has been employed, while adhering to the same evaluation principle.\\ [0.8cm]

PESQ and MOS-LQO  & \(\displaystyle \mathrm{MOS-LQO}=0.999 +\frac{4.999-0.999}{1+e^{-1.4945.PESQ +   4.6607}} \) & \acs{PESQ} serves as an objective technique for evaluating the perceived quality of speech \cite{recommendation2001perceptual}. The assessment involves assigning numerical scores within the range of -0.5 to 4.5. Additionally, a correlation can be established between MOS and \ac{PESQ} scores, giving rise to a novel evaluation metric termed the \ac{MOS-LQO}, also identified as \ac{PESQ} Rec.862.1. \cite{kheddar2022high} \\ [0.8cm]

\acs{RTF}  & \(\displaystyle \mathrm{RTF=\frac{\text{Total Processing Time}}{\text{Total Duration}}}\)  & \Ac{RTF} serves as a standard metric to assess the processing time cost of an \ac{ASR} system. It represents the average processing time required for one second of speech \\[0.6cm]

\acs{RER}  &   \( \displaystyle\frac{(E_{\text{baseline}} - E_{\text{proposed}})}{E_{\text{baseline}}} \times 100\% \) & The  \ac{RER} expresses the percentage error rate achieved by the proposed DL model compared to the baseline. \( E_{\text{baseline}} \) is the error rate of the baseline model. \( E_{\text{proposed}} \) is the error rate of the proposed model or method. \\[0.6cm]

D &  \(\displaystyle \frac{\sum_{i=1}^n \Big( 1 - \frac{\sum_{j=1}^{n} a_{ij} \cdot |i-j|}{\max(|i-1|, |i-2|, \ldots, |i-n|)} \Big) }{n} \hspace{1cm} \)  & Diagonal centrality of an attention matrix (D) is defined as the mean value across the centrality of all its rows. where \(j\) represents the index of each column, \(n\) signifies the length of the input sequence, \(a_{ij}\) denotes the attention weight between the \(i\)-th and \(j\)-th elements of the input sequence, and \(|i-j|\) signifies the distance between the \(i\)-th and \(j\)-th elements of the input sequence \cite{zhang2021usefulness}.\\

 \hline
 \end{tabular}
\end{table*}

\subsection{ASR datasets} 

Various datasets have been employed in the literature for diverse \ac{ASR} tasks. Table \ref{tab:dataset} presents a selection of datasets utilized for \ac{DTL}-based \ac{ASR} applications, along with their respective characteristics. It is important to note that the table primarily includes publicly accessible repositories. Furthermore, it is worth mentioning that certain datasets have undergone multiple updates and improvements over time, leading to their enhanced development.

\begin{table*}[H]
\caption{List of publicly available datasets used for advanced DL-based \ac{ASR} applications}.
\label{tab:dataset}
\scriptsize
\begin{tabular}{m{1.5cm}m{1.5cm}m{4cm}m{9cm}}
\hline
Dataset & Used by & Default ASR task & Characteristics \\
\hline
LibriSpeech & \cite{hrinchuk2020correction,li2021multi,baade2022mae} & Train and assess systems for recognizing speech.  & The collection consists of 1000 hours of speech recorded at a 16 kHz sampling rate, sourced from audiobooks included in the LibriVox project.\\[0.5cm]

DCASE  &\cite{bai2022squeeze,chen2022icnn}& Identifying acoustic environments and detecting sound occurrences. & Comprise 8 coarse-level and 23 fine-level urban sound categories, collected in New York City in 2020 using 50 acoustic sensors.\\[0.5cm]

WSJ &  \cite{zhang2021usefulness} &  Acoustic scene and sound event corpus  &   Comprises an extensive 81 hours of meticulously curated read speech training data. \\[0.5cm]

SWBD &  \cite{zhang2021usefulness}  &  Conversational telephone speech corpus &   Is a comprehensive collection, boasting a substantial 260 hours of conversational telephone speech training data.\\[0.5cm]

AISHELL &  \cite{deng2021improving,zhou2020multi,winata2020lightweight} &  Chinese Mandarin speech corpus &  400 participants from diverse Chinese accent regions recorded in a quiet indoor space using high-fidelity microphones, later downsampled to 16kHz. \\[0.5cm]

CHIME3 & \cite{lee2022regularizing}  & SR for distant microphone in real-world settings.  &  Includes around 342 hours of English speech with noisy transcripts and 50 hours of noisy environment recordings. \\[0.5cm]

Google-SC & \cite{lee2022regularizing}  &  Speech commands with a restricted range of words. &  The dataset contains 105,829 one-second utterances of 35 words categorized by frequency. Each utterance is stored as a one-second WAVE format file with 16-bit single-channel at 16KHz rate. It involves 2,618 speakers \\[0.8cm]

{Aurora-4} & \cite{lee2022regularizing}  & Compare front-ends for large vocabulary recognition performance.  & Aurora-4 is a speech recognition dataset derived from the WSJ corpus, offering four conditions (Clean, channel, noisy, channel+noisy) with two microphone types and six noise types, totaling 4,620 utterances per set.  \\[0.5cm]

Car-env & \cite{lee2022regularizing}    &  Vehicle environment sound &  Is a dataset from Korea that spans 100 hours of recordings in a vehicle. It comprises brief commands, with an average of 1.6 words per command. \\[0.5cm]

HKUST & \cite{li2021multi,winata2020lightweight} & Classify Mandarin speech into standard and accented types. & Comprises roughly 149 hours of telephone conversations in Mandarin.\\[0.5cm]

AudioSet & \cite{baade2022mae}  & Audio event recognition &  Includes 1,789,621 segments of 10 seconds each (equivalent to 4,971 hours). It consists of at least 100 instances clustered into 632 audio classes, with only 485 audio event categories clearly identified.\\[0.5cm]

AWIC-19 & \cite{shareef2022collaborative} & Arabic words recognition & It comprises 770 recordings featuring isolated Arabic words.  \\[0.3cm]

 TED2 &  \cite{fan2023ctc} & English corpus for ASR & The dataset was first made available in May 2012, for training, it comprising 118 hours, 4 minutes, and 48 seconds of training data from 666 speakers, containing approximately 1.7 million words. \\

\hline
\end{tabular}
\end{table*}

\section{Advanced ASR methods and applications} \label{sec3}

 Traditional statistical \acp{LM}, such as backoff n-gram \acp{LM}, have been widely used due to their simplicity and reliability. However, \ac{BERT} \cite{djeffal2023automatic}, which utilize attention models, have shown better contextual understanding compared to single-direction LMs, as demonstrated in the work of Devlin et al. \cite{devlin2019bert}.

In terms of \ac{AM}, deep learning-based models like the deep neural network-hidden Markov model (DNN-HMM) and the \ac{CTC} have made significant advancements \cite{djeffal2023automatic}. DNN-HMM models have been extensively studied in \ac{ASR} research, while \ac{CTC}  is an end-to-end training method that does not require pre-alignment and only needs input and output sequences. The \ac{S2S} model has also been successful in solving \ac{ASR} tasks without using an \ac{LM} or pronunciation dictionary, as described in Chiu et al. \cite{chiu2018state}.

\begin{figure}[ht!]
\centering
\includegraphics[scale=0.65]{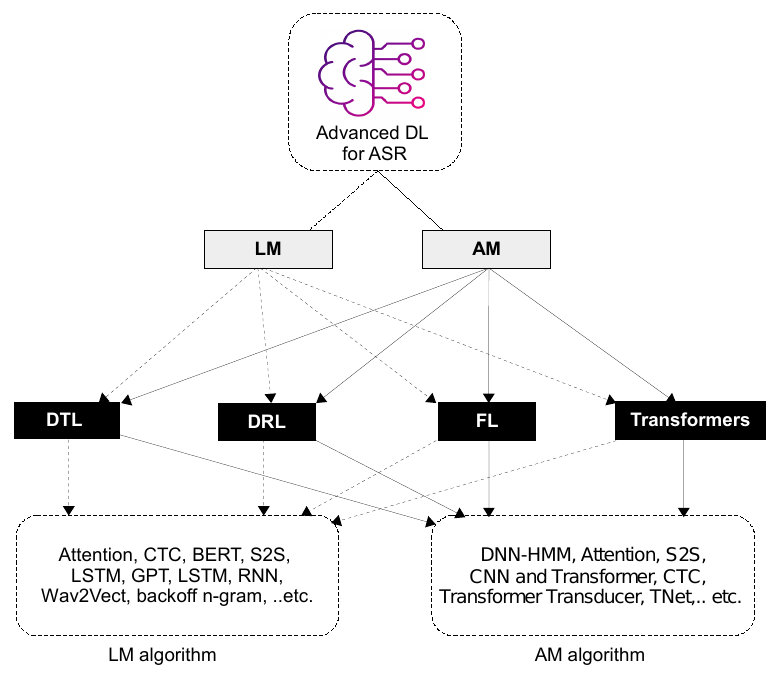}
\caption{Overview of advanced DL-driven \ac{ASR} algorithms and their commonly utilized models.}
\label{fig:04}
\end{figure}

\ac{ASR} systems often face performance degradation in certain situations due to the "one-model-fits-all" approach. Additionally, the lack of diverse and sufficient training data affects \ac{AM} performance. To overcome these constraints and improve the resilience and flexibility of \ac{ASR} systems, advanced DL methodologies such as \ac{DTL} and it sub-field \ac{DA}, \ac{DRL}, and \ac{FL} have surfaced. These innovative methodologies collectively address issues concerning knowledge transfer, model generalization, and training effectiveness, offering remedies that expand upon the capabilities of traditional DL models within the \ac{ASR} sphere. Thus, many research studies have focused on enhancing existing \ac{ASR} systems by applying the aforementioned algorithms. Figure \ref{fig:04} provides an overview of the current \ac{SOTA} advanced DL-based \ac{ASR} and its most useful related schemes in both \ac{AM} and \ac{LM}.

The figures in this section are designed to fulfill two key objectives: firstly, to visually explain the principles behind advanced DL techniques, namely Transformers, \ac{DTL}, \ac{FL}, and \ac{DRL}, as depicted in Figures \ref{fig:ctc}, \ref{domain_adapt}, \ref{fig:fl}, and \ref{fig:rl}, respectively; and secondly, to present practical examples of implementing widely recognized techniques. These include a \ac{CNN}-based Transformer illustrated in Figure \ref{fig:cnnTrans}, a particularly effective \ac{ASR} application of source separation using a Transformer, highlighted in Figure \ref{fig:TRUNet}, and a S2S-based \ac{ASR} model depicted in Figure \ref{fig:SCST}. This dual presentation provides young researchers and developers with valuable, practical insights on how to implement these advanced DL techniques in their projects. By doing so, it bridges the gap between theory and practice, encouraging the application and extension of these examples in new and innovative ways.

\subsection{Transformer-based ASR}

The Transformer stands as a prominent deep learning model extensively employed across diverse domains, including \ac{NLP}, \ac{CV}, and speech processing. Originally conceived for machine translation as a \ac{S2S} model, it has evolved to find applications in various fields. The Transformer heavily relies on the self-attention mechanism, enabling it to capture extensive dependencies in input sequences. The standard Transformer model incorporates the query–key–value (QKV) attention mechanism. In this setup, given matrix representations of queries $\mathbf{Q} \in \mathbb{R}^{N \times D_k}$, keys $\mathbf{K} \in \mathbb{R}^{M \times D_k}$, and values $\mathbf{V} \in \mathbb{R}^{M \times D_v}$, the scaled dot-product attention is defined {as formulated in Equation  \ref{eqAttention}:}

\begin{equation}
 \text{Attention}(\mathbf{Q}, \mathbf{K}, \mathbf{V}) = \text{softmax}\left(\frac{\mathbf{Q}\mathbf{K}^T}{\sqrt{D_k}}\right)\mathbf{V}   
 \label{eqAttention}
\end{equation}

Here, $N$ and $M$ represent the lengths of queries and keys (or values), and $D_k$ and $D_v$ denote the dimensions of keys (or queries) and values. The softmax operation is applied row-wise to the matrix $\mathbf{A}$. Within the Transformer architecture, three attention mechanisms exist based on the source of queries and key–value pairs:

\begin{itemize}
    \item \textbf{Self-attention:} In the Transformer encoder, the queries ($\mathbf{Q}$), keys ($\mathbf{K}$), and values ($\mathbf{V}$) are all equal to the outputs of the previous layer{, considered as $\mathbf{X}$, or to the initial embeddings in the case of the first layer.}  
    \item \textbf{Masked Self-attention: }In the Transformer decoder, self-attention is constrained, allowing queries at each position to attend only to key–value pairs up to and including that position. This is accomplished by implementing a mask function, before normalization,  on the attention matrix  $\hat{\mathbf{A}} = \exp\left(\frac{\mathbf{Q}\mathbf{K}^T}{\sqrt{D_k}}\right)$, where illegal positions are masked out by setting $\hat{A}_{ij} = -\infty$ if $i < j$. This type of self-attention is often referred to as autoregressive or causal attention.

    \item \textbf{Cross-attention:} In cross-attention, queries originate from the results of the preceding (decoder) layer, while keys and values stem from the outputs of the encoder.
\end{itemize}

Numerous studies in the \ac{ASR} field have introduced Transformer-based approaches, encompassing both the acoustic and language domains. In the subsequent subsections, we delve into a comprehensive review and detailed analysis of several cutting-edge techniques within each of these categories. {Table \ref{tab:4} summarises the most recent Transformer-based \ac{ASR} techniques used in \ac{AM} and \ac{LM} domains. }

\begin{table*}[h!]
\scriptsize
\caption{Summary of some proposed work in Transformer-based ASR. The symbol ($\uparrow$) denotes result increase, whereas ($\downarrow$) signifies result decrease. In cases where multiple scenarios are examined, only the top-performing outcome is mentioned.}\label{tab:4}
\begin{tabular}{lm{2.5cm}m{5cm}m{3cm}m{1cm}m{2.5cm}}

\hline
Ref. & Based on  & Speech recognition task  &  Transformer & AM/LM  & Result  with metric \\
 \hline
 \cite{zhou2020multi} &  CNN  &  Solve the problem of code-switching &  Multi-head attention & LM & RER= 10.2\% \\
\cite{winata2020lightweight}  &  VGGnet &  Compress ASR parameters and speeds up the inference time &  Low-rank multi-head attention & AM &  CER= 13.09\% \\
\cite{lee2022regularizing}   &  DNN-HMM  &  Improve ASR &  Attention & AM  & RER= 4.7\%$\downarrow$ \\
\cite{wang2021transformer}  &  Emformer  &  Large scale \ac{ASR} &  Attention & AM &  RERR= 26\%   \\
\cite{aroudi2021trunet}   &  TRUNet  &  Sound source separation &  TNet & AM &  PESQ= 0.22$\uparrow$ \\
\cite{wang2022d}  &  \acs{MHSA}  &  Improve speech/ASR &  D$^2$Net & AM &  PESQ= 0.96$\uparrow$ \\
\cite{shareef2022collaborative}  &  HMM  &  Improve ASR &  Acoustic Encoder & AM & Acc= 96\% \\
\cite{swietojanski2023variable}  &  RNN-T  &  Acoustic re-scoring scenario & Transformer- Transducer & AM  & Acc= 8\%$\uparrow$ \\
\cite{moritz2020all}  &  CTC  &  ASR, ST, Acoustic \ac{ED} &  All-in-one  & AM  & WER=0.3\%$\uparrow$ \\

\cite{huang2020conv}  &  CNN  &  Speech recognition with low latency, reduced frame rate, and streamability. &  Transformer-Transducer & AM & WER= 3.6\%  \\
\cite{fan2021cass}  &  CTC alignment  &  Retrieve the acoustic embedding at the token level for better ASR &  Attention & AM & 51.2x   RTF$\uparrow$ \newline WER= 2.3\% \\

\cite{hadwan2023end}  &  RNN-LSTM  &  Improve the efficiency of end-to-end ASR &  Attention & LM & CER=1.98\% \\
\cite{fan2023ctc}  &  CTC Alignment &  Enhance the performance of  end-to-end \ac{ASR} &  Autoregressive Transformer & AM & RTF= 0.0134  \newline WER= 2.7\% \\
 \hline
\end{tabular}
\end{table*}

\subsubsection{Acoustic domain}

\begin{figure*}
    \centering
    \includegraphics[scale=0.8]{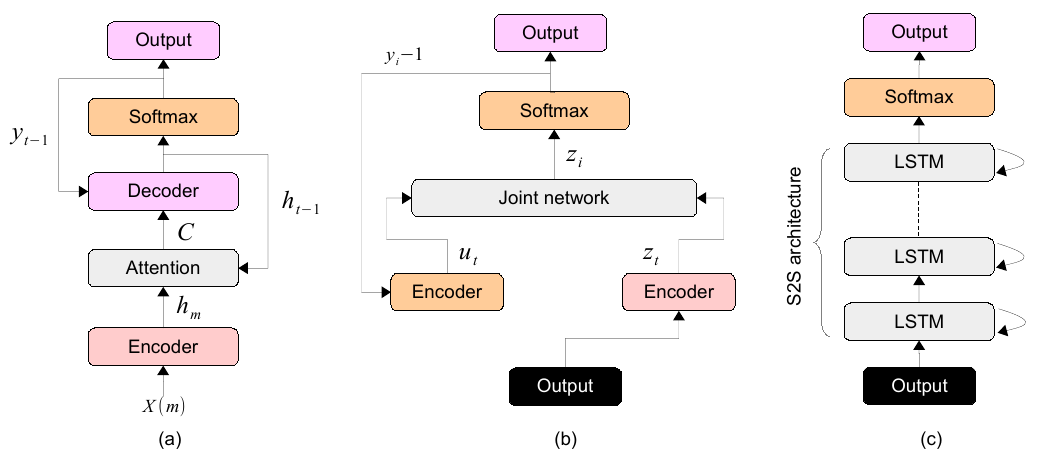}
    \caption{{Three forms of end-to-end Transformers models: (a) attention, (b) RNN-Transducer, and (c) basic CTC  \cite{ahmed2023toward}. }}
    \label{fig:ctc}
\end{figure*}

The study \cite{lee2022regularizing} reveals the Transformer model's increased susceptibility to input sparsity compared to the \ac{CNN}. The authors analyze the performance decline, attributing it to the Transformer's structural characteristics. Additionally, they introduce a novel regularization method to enhance the Transformer's resilience to input sparsity. This method directly regulates attention weights  {, i,e. output of Figure \ref{fig:ctc} (a),} through silence label information in forced-alignment, offering the advantage of not requiring extra module training and excessive computation.

The paper \cite{li2021multi} addresses a limitation in Transformer-based end-to-end modeling for \ac{ASR} tasks, where intermediate features from multiple input streams may lack diversity. The proposed solution introduces a multi-level acoustic feature extraction framework, incorporating shallow and deep streams to capture both traditional features for classification and speaker-invariant deep features for diversity. A \ac{FCF} strategy, employed to combine intermediate features across both the frequency and time domains, correlates and combines these features before feeding them into the Transformer encoder-decoder module.

 The proposed \ac{MAE-AST} operates solely on unmasked tokens \cite{baade2022mae}, utilizing a large encoder. It concatenates mask tokens with encoder output embeddings, feeding them into a shallow decoder. Fine-tuning for downstream tasks involves using only the encoder, eliminating the decoder's reconstruction layers. \ac{MAE-AST} represents a significant improvement over the \ac{SSAST} model for speech and audio classification. Addressing the high masking ratio issue, the method achieves a 3$\times$ speedup and 2$\times$ memory usage reduction. During downstream tasks, the approach consistently outperforms \ac{SSAST}. To identify varities of sounds types, Bai et al. \cite{bai2022squeeze} introduce SE-Trans, a cross-task model for environmental sound recognition, encompassing acoustic scene classification, urban sound tagging, and anomalous sound detection. Utilizing attention mechanisms and Transformer encoder modules, SE-Trans learns channel-wise relationships and temporal dependencies in acoustic features. The model incorporates FMix data augmentation, involving the creation of a binary mask from a randomly sampled complex matrix with a low-pass filter. SE-Trans achieves outstanding performance in \ac{ASR} tasks, proven through evaluations on DCASE challenge databases, underscoring its robustness and versatility in environmental sound recognition.

\begin{figure*}
    \centering
    \includegraphics[scale=0.8]{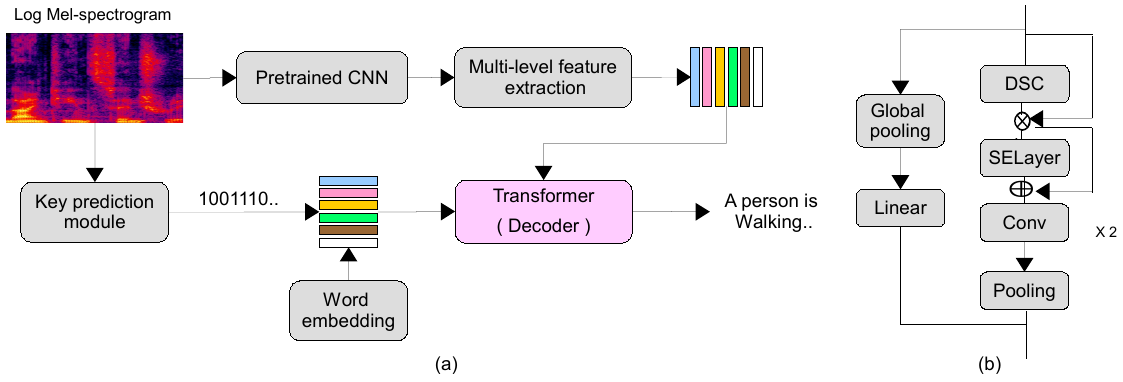}
    \caption{An example of CNN-based Transformer for automated audio captioning \cite{chen2022icnn}.}
    \label{fig:cnnTrans}
\end{figure*}

\Ac{AAC} involves generating textual descriptions for audio recordings, covering sound events, acoustic scenes, and event relationships. Current \ac{AAC} systems typically employ an encoder-decoder architecture, with the decoder crafting captions based on extracted audio features. Chen et al. in their paper \cite{chen2022icnn} introduces a novel approach that enhances caption generation by leveraging multi-level information extracted from the audio clip. The proposed method consists of a CNN encoder with multi-level feature extraction (channel attention, spatial attention), A module specialized in predicting keywords to generate guidance information at the word level and Transformer decoder. Figure \ref{fig:cnnTrans} depicts the overall architecture incorporating the three mentioned modules. Results demonstrate significant improvements in various metrics, achieving \ac{SOTA} performance during the cross-entropy training stage. 

Adversarial audio involves manipulating sound to deceive or compromise \ac{ML} systems, exploiting vulnerabilities in audio recognition models. Both \cite{smietanka2023augmented,li2022adversarial} work are built to combat adversarial noise using Transformers. The authors in \cite{smietanka2023augmented} employed a vision Transformer customized for audio signals to identify speech regions amidst challenging acoustic conditions. To enhance adaptability, they incorporated an augmentation module as an additional head in the Transformer, integrating low-pass and band-pass filters. Experimental results reveal that the augmented vision architecture achieves an F1-score of up to 85.2\% when using a low-pass filter, surpassing the baseline vision Transformer, which attains an F1-score of up to 81.2\%, in speech detection. However, in \cite{li2022adversarial} the authors present an adversarial detection framework using an attention-based Transformer mechanism to identify adversarial audio. Spectrogram features are segmented and integrated with positional information before input into the Transformer encoder, achieving 96.5\% accuracy under diverse conditions such as noisy environments, black-box attacks, and white-box attacks.

The paper \cite{wang2023parallel} introduces a parallel-path Transformer model to address computation cost challenges for speech separation tasks. Using improved feed-forward networks and Transformer modules, it employs a parallel processing strategy with intra-chunk and inter-chunk Transformers. This enables parallel local and global modeling of speech signals, enhancing overall system performance by capturing short and long-term dependencies.

A Hybrid \ac{ASR} approach outlines the conceptualization and execution of a technique that integrates neural network methodologies into advanced continuous speech recognition systems. This integration is built upon \acp{HMM} with the aim of enhancing their overall performance. Wang et al.  \cite{wang2020transformer}  introduce and assess Transformer-based \acp{AM} for hybrid speech recognition. The approach incorporates various positional embedding methods and an iterated loss for training deep Transformers. Demonstrating superior performance on the Librispeech benchmark, the suggested Transformer-based \ac{AM} outperforms the best hybrid result by 19\% to 26\% relative with a standard n-gram \ac{LM}.  

\begin{figure*}
    \centering
    \includegraphics[scale=0.9]{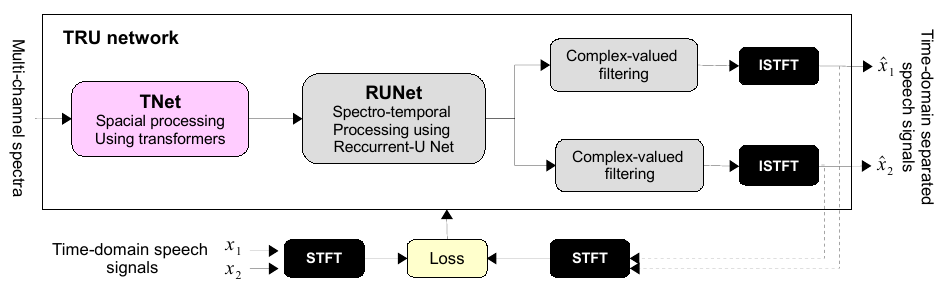}
    \caption{An example of source separation scheme based on Transformer \cite{aroudi2021trunet}.  }
    \label{fig:TRUNet}
\end{figure*}

\Ac{TRUNet} proposed in \cite{aroudi2021trunet} represents an innovative approach to end-to-end multi-channel reverberant sound source separation. The model incorporates a recurrent-U network that directly estimates multi-channel filters from input spectra, enabling the exploitation of spatial and spectro-temporal diversity in sound source separation. In Figure \ref{fig:TRUNet}, the block diagram illustrates the architecture of \ac{TRUNet}'s Transformer. The \ac{TNet} encompasses three variations: (i) TNet–Cat, which concatenates multi-channel spectra, treating them as a single input. This approach allows for the direct utilization of spatial information between channels. (ii) TNet–RealImag, utilizing two separate Transformer stacks for real and imaginary parts, respectively. Queries and keys are computed from the multi-channel spectra. Despite this, the method may not fully exploit spatial information, such as phase differences, directly. (iii) TNet–MagPhase, analogous to TNet–RealImag, but employing spectral magnitude and spectral phase instead of real and imaginary parts. This variation proves superior in extracting spatial information from complex-valued spectra, resulting in maximum enhanced performance in sound source separation when employing TRUNet-MagPhase architecture. \Ac{STFT}, and \ac{ISTFT}, is to analyze and reconstruct signals in the time-frequency domain, respectively.

In recent times, dual-path networks have demonstrated effective results in many speech processing tasks such as speech separation and \ac{SE}. In light of this, Wang Ke and colleagues incorporated the Transformer into the structure of dual-path networks, presenting a time-domain \ac{SE} model called two-stage \ac{TSTNN}. This model significantly enhances the performance of \ac{SE}  \cite{wang2021tstnn}. Some research findings suggest that the dot-product self-attention may not be essential for Transformer models. Similarly, the paper \cite{wang2022d} introduces D$^2$Net, a denoising and dereverberation network for challenging single-channel mixture speech in complex acoustic environments. D$^2$Net incorporates a \ac{TBE} for feature extraction and fusion, along with a \ac{GLDPT} featuring local dense synthesizer attention to enhance local information perception.

The study proposed by {Gong et al. } \cite{gong2022ssast} explores self-attention-based neural networks like the \ac{AST} for audio tasks. It introduces a self-supervised framework, improving \ac{AST} performance by 60.9\% on various speech classification tasks such as \ac{ASR}, speaker recognition, and more, reducing reliance on labeled data. The approach marks a pioneering effort in audio self-supervised learning. Likewise, a novel augmented memory self-attention addresses limitations of Transformer-based acoustic modeling in streaming applications has been proposed in \cite{wu2020streaming}, outperforming existing streamable methods by over 15\% in relative error reduction on benchmark datasets.

Shareef et al. in  \cite{shareef2022collaborative} propose a collaborative training method for acoustic encoders in Arabic \ac{DTL} systems for speech-impaired children, achieving a 10\% relative accuracy improvement on phoneme alignment in the output sequence. Pioneering in recognizing impaired children's Arabic speech. Similarly in \cite{nagaraja2021collaborative}, collaboratively  training acoustic encoders of various sizes for on-device \ac{ASR} improves efficiency and reduces redundancy. Using co-distillation with an auxiliary task, collaborative training achieves up to 11\% relative \ac{WER} improvement on LibriSpeech corpus. 

Transducer models {(Figure \ref{fig:ctc} (b))}, in the context of \ac{ASR}, map input sequences (acoustic features) to output sequences (transcriptions). Unlike traditional \ac{S2S} models, transducers can handle variable-length input and output sequences more efficiently. The study \cite{swietojanski2023variable} explores attention masking in Transformer-Transducer-based \ac{ASR}, comparing fixed masking with chunked masking in terms of accuracy and latency. The authors claim that variable masking is the viable choice in acoustic rescoring scenarios. Similarly, to adapt the Transformer for streaming \ac{DTL}, the authors in \cite{huang2020conv} employ the Transducer framework for streamable alignments. Using a unidirectional Transformer with interleaved convolution layers for audio encoding, they model future context and gradually downsample input to reduce computation cost, while limiting history context length.

Moving on, the work \cite{moritz2020all} introduces an all-in-one \ac{AM} based on the Transformer architecture, combined with the \ac{CTC} to ensure a sequential arrangement and utilize timing details. It addresses \ac{ASR}, \ac{AT}, and acoustic \ac{ED} simultaneously. The model demonstrates superior performance, showcasing its suitability for comprehensive acoustic scene transcription. Winata et al. \cite{winata2020lightweight} propose a memory-efficient Transformer architecture for end-to-end speech recognition. It significantly reduces parameters, boosting training speed by over 50\% and inference time by 1.35$\times$ compared to baseline. Experiments show better generalization, lower error rates, and outperformance of existing schemes without external language or acoustic models. Growing demand for on-device \ac{ASR} systems prompts interest in model compression.

\subsubsection{Language domain}

Self-attention models, such as Transformers, excel in speech recognition and reveal an important pattern. As upper self-attention layers are replaced with feed-forward layers, resembling \ac{CLDNN} architecture in \cite{zhang2021usefulness}, experiments on \ac{WSJ} and \ac{SWBD} datasets show no performance drop and minor gains. The novel proposed metric of attention matrix diagonality indicates increased diagonality in lower to upper encoder self-attention layers. The authors conclude that a global view appears unnecessary for training upper encoder layers in speech recognition Transformers when lower layers capture sufficient contextual information. The study conducted by Hrinchuk et al. \cite{hrinchuk2020correction} presents a proficient postprocessing model for \ac{ASR} with a Transformer-based encoder-decoder architecture, initialized with the weights of pre-trained \ac{BERT} model \cite{Kheddar023ASR}. The model effectively refines \ac{ASR} output, demonstrating substantial performance gains through strategies like extensive data augmentation and pretrained weight initialization. On the LibriSpeech benchmark dataset, significant reductions in \acp{WER} are observed, particularly on noisier evaluation dataset portions, outperforming   baseline models and approaching the performance of Transformer-XL neural language model re-scoring with 6-gram.

\Ac{CTC} is an architecture and principle commonly used in \ac{S2S} tasks {(Figure \ref{fig:ctc} (c))}, such as \ac{ASR}. It enables alignment-free training by introducing a blank symbol and allowing variable-length alignments between input and output sequences. During training, the model learns to align the input sequence with the target sequence, and the blank symbol accounts for multiple possible alignments. \ac{CTC} is particularly effective in tasks with variable-length outputs, making it well-suited for applications like speech recognition where the duration of spoken words may vary. This latter has been used in many \ac{ASR} schemes, for example, Deng et al.   \cite{deng2021improving} presents the innovative pretrained Transformer \ac{S2S} \ac{ASR} architecture, which integrates self-supervised pretraining techniques for comprehensive end-to-end \ac{ASR}. Employing a hybrid \ac{CTC}/attention  model, it maximizes the potential of pretrained \ac{AM} and \ac{LM}. The inclusion of a \ac{CTC} branch aids in the encoder's convergence during training and considers all potential time boundaries in beam searching. The encoder is initiated with wav2vec2.0, and the introduction of a one-cross decoder mitigates reliance on acoustic representations, enabling initialization with pretrained DistilGPT2 and overcoming the constraint of conditioning on acoustic features.

\Ac{CS} takes place when a speaker switches between words of two or more languages within a single sentence or across sentences. Zhou et al. \cite{zhou2020multi} introduces a multi-encoder-decoder Transformer, for \ac{CS} problem. It employs language-specific encoders and attention mechanisms to enhance acoustic representations, pre-trained on monolingual data to address limited \ac{CS} training data.
Hadwan et al. research \cite{hadwan2023end} employ an attention-based encoder-decoder Transformer, to enhance end-to-end \ac{ASR} for the Arabic language, focusing on Qur'an recitation. The proposed model incorporates a multi-head attention mechanism and Mel filter bank for feature extraction. For constructing a \ac{LM}, \ac{RNN} and \ac{LSTM} techniques were employed to train an n-gram word-based \ac{LM}. The study introduces a new dataset, yielding \ac{SOTA} results with a low character error rate.

\subsection{{DTL}-based ASR}

Overall, \ac{DTL} consists of training a DL model on a specific domain (or task) and then transferring the acquired knowledge to a new, similar domain (or task). In what follows, we present some of the definitions that are essential to understand the principle of \ac{DTL} for \ac{ASR} applications.

\begin{figure}[t!]
\begin{center}
\includegraphics[width=0.45\textwidth]{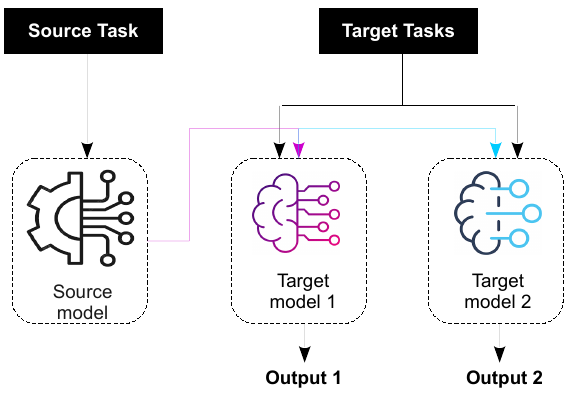}\\
\end{center}
\caption{Deep transfer learning principle.}
\label{domain_adapt}
\end{figure}

\Ac{DTL} refers to a \ac{DL} paradigm where knowledge gained from pre-training a model (Source model) on one domain or task is leveraged to enhance performance of target model, on a different but related domain or task. In this context, a "domain" refers to a specific data distribution, while a "task" represents a learning objective. \textit{\Ac{DA}} in \ac{DTL} involves adapting a model trained on a $D_S$ to perform well on a target domain. This is crucial when there are differences in data distributions between the two domains. \textit{Fine-tuning} is a technique where a pre-trained model is further trained on task-specific data to improve its performance on a related task. \textit{Cross-domain learning} extends transfer learning to scenarios where the source and target domains are distinct. \textit{Zero-shot learning} involves training a model to recognize classes not present in the training data. \textit{Transductive} \ac{DTL} focuses on adapting a model based on a specific set of target instances.  \textit{Inductive} \ac{DTL} aims to generalize knowledge across domains by training a model to handle diverse tasks and domains simultaneously \cite{Kheddar023ASR,himeur2023video,kheddar2023deep}. These techniques contribute to the versatility and adaptability of \ac{DL} models in various applications. {Figure } \ref{domain_adapt} depicted the principle of \ac{DTL} techniques. Table \ref{tab:5} summarises the most recent DTL-based \ac{ASR} techniques used in \ac{AM} and \ac{LM} domains.

\subsubsection{Acoustic domain}
Schneider et al. \cite{schneider2019wav2vec} explored unsupervised pre-training for speech recognition using wav2vec model on large unlabeled audio data. The learned representations enhanced \ac{AM} training with a simple \ac{CNN} optimized through noise contrastive binary classification.
In \cite{thienpondt2022transfer}, a source filter warping data augmentation strategy is proposed to enhance the robustness of children's speech \ac{ASR}. The authors constructed an end-to-end acoustic model using the XLS-R wav2vec 2.0 model, pre-trained in a self-supervised manner on extensive cross-lingual corpora of adult speech. The work proposed in \cite{dan2022multi} introduces a multi-dialect acoustic model employing soft-parameter-sharing multi-task learning, a transductive \ac{DTL} subcategory, within the Transformer architecture. Auxiliary cross-attentions aid dialect ID recognition, providing dialect information. Adaptive cross-entropy loss automatically balances multi-task learning. Experimental results demonstrate a significant reduction in error rates compared to various single- and multi-task models on multi-dialect speech recognition and dialect ID recognition tasks. Similarly, in the realm of computer vision, \acp{CNN} models like ConvNeXt have outperformed cutting-edge Transformers, partly due to the integration of \ac{DSC}. \ac{DSC}, which approximates regular convolutions, enhances the efficiency of \acp{CNN} in terms of time and memory usage without compromising accuracy—in some cases, even enhancing it. The study \cite{pellegrini2023adapting}   introduces \ac{DSC} into the pre-trained audio model family for audio classification on AudioSet (target task), demonstrating its advantages in balancing accuracy and model size. Xin et al. \cite{xin2022audio} introduce an audio pyramid Transformer with an attention tree structure, with four branches, to reduce computational complexity in fine-grained audio spectrogram processing. It proposes a \ac{DA} transfer learning approach for weakly supervised sound \ac{ED}, a sub-field of \ac{ASR}, enhancing localization performance by aligning feature distributions between frame and clip domains with a \ac{DA} detection loss.

\subsubsection{Language domain}

The methodology is founded on the utilization of the \ac{BERT} model \cite{devlin2018bert}, which involves pretraining language models and demonstrates improved performance across various downstream tasks.
\ac{DTL} approaches for language models, specifically employed in the domain of voice recognition, are referred to as \ac{LM} adaptation. These approaches aim to bridge the gap between the source distribution $\mathbb{D}_{S}$ and the target distribution $\mathbb{D}_{T}$.
Song et al. \cite{song2019l2rs} present L2RS approach, which relies on two main components: (i) utilizing diverse textual data from \ac{SOTA} \ac{NLP} models, such as \ac{BERT}, and (ii) automatically determining their weights to rescore the N-best lists for \ac{ASR} systems.

Recent advancements in \ac{S2S} models have shown promising results for training monolingual \ac{ASR} systems. The \ac{CTC} and encoder-decoder models are two popular architectures for end-to-end \ac{ASR}. Additionally, joint training of these architectures in a multi-task hybrid approach has been explored, demonstrating improved overall performance. For instance, the architecture illustrated in {Figure } \ref{fig:ctc} (a) comprises \ac{S2S} layers. The encoder network of \ac{S2S} consists of a series of \acp{RNN} that generate embedding vectors, while the \ac{RNN} decoder utilizes these vectors to produce final results. The \ac{RNN} also benefits from prior predictions ($P_i,  i=0,\dots, n$), enhancing the accuracy of subsequent predictions. Moving on, a novel DTL-based approach that enhances end-to-end speech recognition has been proposed in \cite{qin2018towards}. 
The novelty lies in applying \ac{DTL} through multilingual training and multi-task learning at two levels. The initial stage utilizes non-negative matrix factorization, instead of a bottleneck layer, and multilingual training for high-level feature extraction. The subsequent stage employs joint CTC-attention models on these features, where the \ac{CTC} was transferred to the target attention-based model. The scheme demonstrated superior performance on TIMIT but requires testing on high-resource data. Further optimization is needed for standard end-to-end training. In addition, integrating both \ac{AM} and \ac{LM} methodologies has the potential to enhance or construct an effective \ac{DTL}-based \ac{DTL} model, as demonstrated in \cite{bai2021fast,jiang2021gdpr,weninger2019listen,deena2018recurrent,ng2020cuhk,manohar2023automatic}. 

\begin{table*}[h!]
\caption{In contemporary cutting-edge frameworks, diverse pre-trained models are utilized for distinct tasks within the field. These frameworks employ different \ac{DTL} approaches and assess their efficacy using specific metrics. The symbol ($\uparrow$)  result increase, whereas ($\downarrow$) signifies result decrease. In cases where multiple scenarios are examined, only the top-performing outcome is mentioned.}
\label{tab:5}
\scriptsize
\begin{tabular}{lp{2.5cm}p{5cm}lp{2.5cm}p{2.8cm}}
\hline
Scheme & Based on & Speech recognition task  ($\mathbb{T}_{T}$)  & AM/LM & Adaptation & Result with metric \\
 \hline

\cite{pellegrini2023adapting} & ConvNeXt-Tiny & Audio classification & AM & DA & \acs{mAP}= 0.471 \\

\cite{xin2022audio} & \acs{APT} & Sound event detection & AM & DA & F1= 79.6\% \\

\cite{hrinchuk2020correction}  &  BERT (Jasper) &  Speech-to-text & LM & DTL & WER= 14\% \\

\cite{deng2021improving} &   DistilGPT2  &  Improve ASR & Both & Fine-tuning & CER= 4.6\% \\

\cite{manohar2023automatic}& XLRS Wave2vec&Improve ASR in low resource language & Both & Fine-tuning & 5.6\% WER $\downarrow$\\

\cite{thienpondt2022transfer} & XLRS Wave2vec &Improve ASR for children's speech & AM & DTL &  WER = 4.86\%\\

\cite{weninger2019listen}  & \ac{S2S} & Speaker adaptation  & Both & Features norm.  & 25.0\% WER$\downarrow$\\

\cite{dan2022multi} & Transformer &  Multi-dialect model aids recognizing diverse speech dialects effectively & AM & Multi-task learning & Acc= 100\%\\

\cite{cho2018multilingual}& \acs{S2S} & Enhancing the existing multilingual \ac{S2S} model. &LM&DTL&4\%CER  $\downarrow$ 6\% WER \\

\cite{lifine} &  PaSST  &  Audio tagging and event detection  &  AM  & Fine-tuning &  F1= 64.85\% \\

\cite{schneider2019wav2vec}&Wav2vec&WSJ data speech &AM&affine transform.& 36\% WER$\downarrow$\\

\cite{wang2022arobert}&ARoBERT& Spoken language understanding &LM&Fine-tuning&F1-score=92.56\%\\

 \hline
\end{tabular}
\begin{flushleft}
 Abbreviations: Transformer (T)   
\end{flushleft}
\end{table*}

\subsection{FL-based ASR}

\Ac{FL} revolutionizes \ac{AI} model training by enabling collaboration without the need to share sensitive training data. Traditional centralized approaches are evolving towards decentralized models, where \ac{ML} algorithms are trained collaboratively on edge devices like mobile phones, laptops, or private servers \cite{himeur2023federated}. The mathematical formulation of \ac{FL} focuses on training a single global model across multiple devices or nodes (clients) while keeping the data localized. The objective is to minimize a global loss function that is typically the weighted sum of the local loss functions on all clients. The standard \ac{FL} problem can be formulated as:
\begin{equation}
\label{FL}
\min_{\theta} F(\theta) = \min_{\theta} \sum_{k=1}^{K} \frac{n_k}{N} F_k(\theta) 
\end{equation}

In this context, $ \theta $ denotes the parameters of the global model to be learned, $ K $ represents the total number of clients, $ n_k $ signifies the number of data samples at client $ k $, $ N = \sum_{k=1}^{K} n_k $ stands for the total number of data samples across all clients, and $ F_k(\theta) $ indicates the local loss function computed on the data of client $ k $. In \ac{FL}, the goal is to find the global model parameters $ \theta $ that minimize the global loss function $ F(\theta) $, which is an aggregate of the local loss functions from all participating clients. This process typically involves iterative updates to the model parameters using algorithms like \ac{FedAvg}, where clients compute gradients or updates based on their local data and send these updates to a central server. The server then aggregates these updates to improve the global model.

\begin{itemize}
    \item \textbf{Horizontal federated {learning} (HFL)}:
In \acs{HFL}, clients train a shared global model using their respective datasets, characterized by the same feature space but different sample spaces. Each client utilizes a local \ac{AI} model, and their updates are aggregated by a central server without exposing raw data. The \acs{HFL} training process involves: (1) initialization, (2) local training, (3) encryption of gradients, (4) secure aggregation, and (5) global model parameter updates. The objective function minimizes a global loss across all parties' datasets \cite{himeur2023federated}.

\item \textbf{\Ac{VFL}:}
\ac{VFL} trains models on datasets sharing the same sample space but having different feature spaces. Through entity data alignment (EDA) and  encrypted model trained (EMT), \ac{VFL} allows clients to cooperatively train models without sharing raw data. The training process involves the same steps as \acs{HFL} \cite{himeur2023federated}. 
\end{itemize}

\begin{figure*}
    \centering
    \includegraphics[scale=0.65]{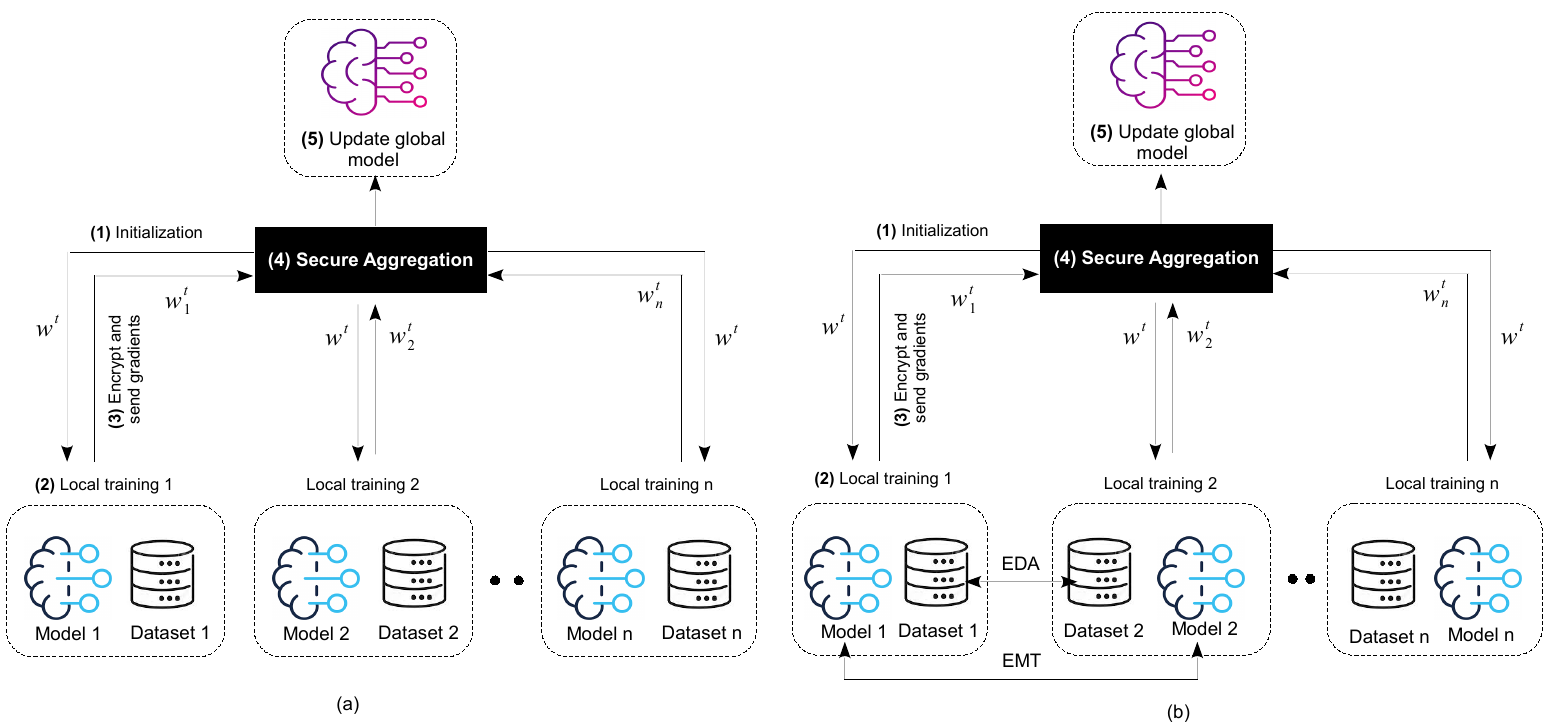}
    \caption{Federated learning principle: (a) Horizontal FL, (b) Vertical FL.}
    \label{fig:fl}
\end{figure*}

FL presents a paradigm shift in \ac{AI} training, promoting collaboration while respecting data privacy. The working principle of \acs{HFL}, and \ac{VFL} is depicted on Figure \ref{fig:fl}, they cater to various data distribution scenarios, offering flexible solutions for decentralized and secure \ac{ML}. The application of these FL frameworks extends across diverse domains, promising improved model accuracy and privacy preservation. The first work introducing \ac{FL} in \ac{ASR} is presented in 
\cite{dimitriadis2020federated}. The authors introduced a FL platform that is easily generalizable, incorporating hierarchical optimization and a gradient selection algorithm to enhance training time and SR performance. Guliani et al.  \cite{guliani2021training} 
proposed a strategy to compensate
\ac{non-IID} data in federated training of \ac{ASR}
systems. The proposed strategy involved random client data sampling, which resulted
in a cost-quality trade-off.  
Zhu et al. \cite{zhu2022decoupled} addressed also FL-based \ac{ASR} in \ac{non-IID} scenarios with personalized \ac{FL}. They introduced two approaches: adapting personalization layer-based FL for \ac{ASR}, involving local layers for personalized models, and proposing decoupled federated learning (DecoupleFL). DecoupleFL reduces computation on clients by shifting the computation burden to the server. Additionally, it communicates secure high-level features instead of model parameters, reducing communication costs, particularly for large models.
In \cite{cui2021federated}, the authors proposed a
client-adaptive federated training scheme to mitigate data heterogeneity when training
\ac{ASR} models. Nguyen et al. \cite{nguyen2023federated} used FL to train an \ac{ASR} model based on a wav2vec 2.0 model pre-trained by self supervision. Yang et al. \cite{yang2021decentralizing} proposed a  decentralized feature extraction approach in
federated learning. This approach  is built upon a \ac{QCNN} composed of a quantum circuit encoder for feature extraction, and an RNN based end-to-end \ac{AM}. This framework takes advantage of the quantum learning progress to secure models and to avoid privacy
leakage attacks. Gao et al. \cite{gao2022end} tackled a challenging and realistic \ac{ASR} federated experimental setup with clients having heterogeneous data distributions, featuring thousands of different speakers, acoustic environments, and noises. Their empirical study focused on attention-based \ac{S2S} end-to-end \ac{ASR} models, evaluating three aggregation weighting strategies: standard \ac{FedAvg}, loss-based aggregation, and a novel WER-based aggregation. 
Table \ref{tab:fl} summarizes the most recent FL-based \ac{ASR} techniques. 


\begin{table*}[h!]
\scriptsize
\caption{Summary of recent proposed work in FL-based \ac{ASR}. All the schemes are suggested for \ac{AM}. The symbol ($\uparrow$)  result increase, whereas ($\downarrow$) signifies result decrease. In cases where multiple scenarios are examined, only the top-performing outcome is mentioned.}
\label{tab:fl}
\begin{tabular}{lm{4cm}m{4cm}m{3cm}m{4cm}}

\hline
Scheme &Based on  & Speech recognition task  &  FL technique  & Metric and result\\
 \hline
\cite{dimitriadis2020federated}  &  \ac{S2S}  &  Improve ASR  &  FedAvg &  WER = 6\% \\ 

\cite{guliani2021training}  &  End-to-end RNN-T &  ASR on non-IDD data &  FedAvg &    WER= \\ 

\cite{zhu2022decoupled}  &  CNN+Transformer extractor  &  ASR on non-IDD data &  DecoupleFL &   2.3- 3.4\% WER $\downarrow$ compared with FedAvg  \\ 

\cite{cui2021federated}  &  LSTM  &  ASR on non-IDD data &  \acs{CAFT} &   WER = 15.13\% \\ 

\cite{nguyen2023federated}  &  wav2vec 2.0  &  Improve ASR &  FedAvg & WER= 10.92\% \newline EER= 5-20\%  \\ 

\cite{yang2021decentralizing}  &  QCNN  and RNN   & Improve privacy-preservation in ASR &  VFL&  Accuracy = 95.12\% \\ 

\cite{gao2022end}  &  \ac{S2S} & ASR on heterogeneous data distributions &  FedAvg &   WER= 19.98-23.86\% \\ 

\cite{mehmood2022fednst}  &  \ac{S2S}  & ASR on private and unlabelled
user data. &  \acs{FedNST} &   WER= 22.5\% \\ 

\cite{vasquez2023novel}  &  Wav2vec 2.0 and Whisper  &  ASR \& KWS for child exploitation settings &  FedAvg  &  WER = 11-25\%\\ 

\cite{tan2020federated}  &  Kaldi and backoff n-gram  &  Improve privacy-preservation in ASR & Merging models &  WER= 17.7\% \\ 

\cite{tomashenko2022privacy}  &  TDNN  &  Improve privacy-preservation in ASR &  Aggregation &  EER = 1-2\%. \\ 

\cite{guliani2022enabling}  &  Non-Streaming \& Streaming Conformer  &  Reduce client ASR model size  &  Federated Dropout  & 6-22\% $\downarrow$ Client size model; 34-3\% WER $\downarrow$   \\

 \hline
\end{tabular}
\end{table*}

\subsection{DRL-based ASR}

\begin{figure}
    \centering
    \includegraphics[scale=0.85]{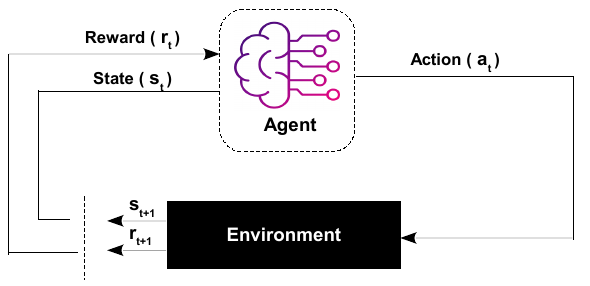}
    \caption{DRL principle.}
    \label{fig:rl}
\end{figure}

\Ac{DRL} is a \ac{ML} paradigm where an agent learns optimal decision-making by interacting with an environment. The agent receives feedback in the form of rewards or penalties, adapting its behavior to maximize cumulative reward over time through a trial-and-error process. \ac{DRL} involves several key concepts, as defined in the following terms: Environment model, serving as a representation of contextual dynamics; State ($s$), denoting the current situation perceived by the agent; Observation ($o$), a subset of the state directly perceived by the agent; Action ($a$), the decision made by the agent in response to the environment; Policy ($\pi$), describing how the agent converts environmental conditions into actions; Agent, the entity making decisions based on current states and past experiences; Reward, numerical values assigned by the environment to the agent based on state-action interactions. Figure \ref{fig:rl} illustrates the principle of \ac{DRL}.

\ac{MDP} is a fundamental framework for dynamic and stochastic decision-making, characterized by state space \(S\), action space \(\mathbb{A}\), transition probabilities \(\mathbb{P}\), and a reward function \(R\). The primary objective in an \ac{MDP} is to identify an optimal policy \(\pi^*\) maximizing the expected discounted total reward over time, expressed as:

\begin{equation}
   \pi^* = \max_{\pi} \mathbb{E}_{\pi}[\sum_{t=0}^T \gamma^t r_t(s_t, \pi(s_t))], 
\end{equation}

\noindent where \(\gamma\) is the discount factor. \acp{MDP} find extensive applications in addressing uncertainties in intelligent systems within dynamic wireless environments, including spectrum management, cognitive radios, and wireless security.

\begin{figure}
    \centering
    \includegraphics[scale=0.8]{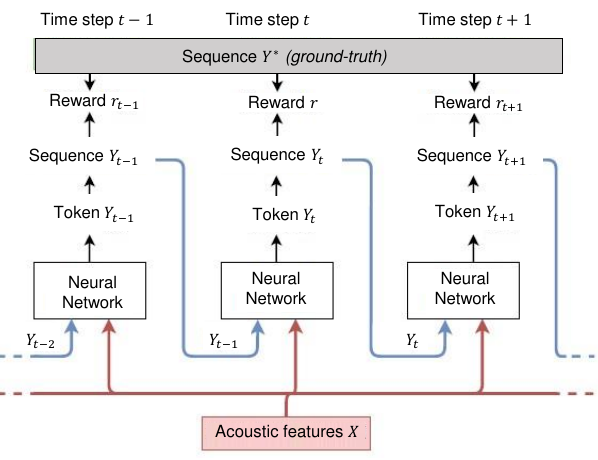}
    \caption{Example of sequential decision model of DRL-based \ac{ASR} \cite{chen2022self}.}
    \label{fig:SCST}
\end{figure}

In the field of \ac{ASR}, \ac{DRL} has primarily been proposed to tackle discrepancies between the training and testing phases. Two main discrepancies leading to potential performance deterioration have been identified: 1) The conventional use of the cross-entropy criterion maximizes log-likelihood during training, while performance is assessed by \ac{WER}, not log-likelihood; 2) The teacher-forcing method, which relies on ground truth during training, implies that the model has never encountered its own predictions before testing. \ac{DRL} addresses these discrepancies by bridging the gap between the training and testing phases. Several \ac{DRL}-based approaches for \ac{ASR} have been proposed \cite{chen2022self, kala2018reinforcement, tjandra2018sequence, tjandra2019end, dudziak2019shrinkml, mehrotra2020iterative, shen2019reinforcement, tsai2019using, chung2020semi, chen2023end}. For example, in \cite{chen2022self}, the authors introduced a \ac{DRL}-based optimization method for the \ac{S2S} \ac{ASR} task called \ac{SCST}. This method can be conceptualized as a sequential decision model, depicted in Figure \ref{fig:SCST}. The entire encoder-decoder neural network is treated as an agent. At each time step \(t\), the current state \(s_t\) is formed by concatenating the acoustic feature \(x_t\) and the previous prediction \(Y_{t-1}\). The output token serves as the action, updating the generated hypotheses sequence. \ac{SCST} associates training loss and \ac{WER} using a WER-related reward function, calculating the reward \(r_t\) at each token generation step by comparing it with the ground truth sequence \(Y^*\). \ac{SCST} uses the test-time beam search algorithm to sample hypotheses for reward normalization, assigning positive weights to high-reward hypotheses that outperform the current test-time system and negative weights to low-reward hypotheses. The framework is illustrated in Figure \ref{fig:SCST}.

In \cite{kala2018reinforcement}, the authors developed a \ac{DRL} framework for speech recognition systems using the policy gradient method. They introduced a \ac{DRL} method within this framework, incorporating user feedback through hypothesis selection.
Tjandra et al. \cite{tjandra2018sequence, tjandra2019end} also employed policy gradient \ac{DRL} to train a \ac{S2S} \ac{ASR} model.
In \cite{dudziak2019shrinkml}, the authors constructed a generic \ac{DRL}-based AutoML system. This system automatically optimizes per-layer compression ratios for a \ac{SOTA} attention-based end-to-end \ac{ASR} model, which consists of multiple \ac{LSTM} layers. The compression method employed in this work is \ac{SVD} low-rank matrix factorization.  The authors improved  this approach by combining iterative compression with AutoML-based rank searching, achieving over 5 x \ac{ASR} compression without degrading the WER \cite{mehrotra2020iterative}.  
Shen et al. \cite{shen2019reinforcement} suggested employing \ac{DRL} to optimize a \ac{SE} model based on recognition results, aiming to directly enhance \ac{ASR} performance.  AutoML-based \ac{LRF} achieves up to 3.7× speedup. In the shade of this, Mehrotra et al. in their work \cite{mehrotra2020iterative} propose an iterative AutoML-based \ac{LRF} that employs \ac{DRL} for the iterative search, surpassing 5× compression without degrading \acp{WER}, advancing \ac{ASR}.  Table \ref{tab:rl} summarizes the recent \ac{DRL}-based \ac{ASR} techniques.  
 
\begin{table*}[h!]
\scriptsize
\caption{Summary of recent proposed works in \ac{DRL}-based ASR. The symbol ($\uparrow$)  result increase, whereas ($\downarrow$) signifies result decrease. In cases where multiple scenarios are examined, only the top-performing outcome is mentioned.}
\label{tab:rl}
\begin{tabular}{lm{3cm}m{4cm}m{2.5cm}m{4cm}}

\hline
Scheme & Model-based   & ASR Tasks  &  DRL technique & Metric and result\\ \hline

\cite{chen2022self} & \ac{S2S} conformer  & Improve ASR  &  Policy gradient &  8.7\%-7.8\% WER $\downarrow$ over Baseline model \\ 
\cite{kala2018reinforcement} &  DNN-HMM  & Improve ASR for AM  &  Policy gradient  &  WER=23.82-25.43\%  \\ 
\cite{tjandra2018sequence} &  \ac{S2S}  & Improve ASR  for AM &  Policy gradient  &  CER=6.10\%  \\ 
\cite{tjandra2019end} &   \ac{S2S}  &  Improve ASR  for AM &  Policy gradient  &  CER=6.10\% \\ 
\cite{dudziak2019shrinkml} &   End-to-end encoder-attention-
decoder &  Improve ASR Model compression for AM &  Policy gradient  &  Up to \(\sim\)3x compression; WER=8.06\% \\ 
\cite{mehrotra2020iterative} &  End-to-end encoder-attention-
decoder  & Improve ASR Model compression for AM&  Policy gradient   & Up to \(\sim\)5x compression; WER=8.19\% \\ 

\cite{shen2019reinforcement} & CD-DNN-HMM AM \& SRI LM &  Speech enhancement for ASR for AM and LM & Q-learning    & 12.40\% and 19.23\% CER $\downarrow$ at 5 and 0 dB SNR conditions. \\ 
\cite{tsai2019using} &  LSTM  & Improve ASR for dialogue state tracking  &  DQN  &  Acc= 3.1\%$\uparrow$ \\ 
\cite{chung2020semi} & \ac{S2S}   & Improve ASR for AM & Policy gradient    &   CER=8.7\%  \\ 
\cite{chen2023end} & Wav2vec 2.0   & Improve ASR for AM &  Policy gradient  &     4\% WER $\downarrow$ \\

\hline
\end{tabular}
\end{table*}

\section{Open Issues and Key challenges} \label{sec4}

Integrating advanced techniques like \ac{DTL}, \ac{FL}, and \ac{DRL} into \ac{ASR} systems presents exciting opportunities but comes with its set of challenges. This section delves into the distinct challenges associated with each approach, emphasizing the critical areas that demand attention and innovation.

\subsection{Transformer-based}

Transformer-based \ac{ASR} systems have revolutionized the field of speech processing with their superior ability to handle sequential data. However, their deployment in the acoustic and language domains presents unique challenges and open issues. \textbf{\textit{For the acoustic domain}}, one of the key challenges in the \ac{AM} is the handling of long audio sequences. Transformers require significant memory and computational resources, making it difficult to process lengthy audio inputs efficiently. Additionally, the variability in speech, such as accents, speech disorders, and background noise, can affect the robustness of transformer-based \ac{ASR} systems. Developing models that can generalize across these variations without substantial data for each scenario remains an open issue. \textbf{\textit{For the language domain}}, the major challenge  in the \ac{LM} lies in capturing the nuances of different languages and dialects. The transformer's reliance on large amounts of training data poses a challenge for low-resource languages. Furthermore, language models need to understand context deeply to accurately predict words in continuous speech, which can be particularly challenging in languages with rich morphology. \textbf{\textit{For both domains}}, integrating acoustic and language models in transformer-based \ac{ASR} systems to work seamlessly is complex. Achieving real-time processing speeds while maintaining high accuracy is a persistent challenge. Furthermore, the interpretability of these models is limited, making it difficult to diagnose and correct errors in \ac{ASR} predictions. Addressing the trade-off between model complexity and the ability to deploy these systems on devices with limited computing power is also a critical challenge.

\subsection{DTL and DA-based}

This section discusses challenges and concepts related to \ac{DTL} and \ac{DA} in speech recognition, including distribution shift, feature space adaptation, label distribution shift, catastrophic forgetting, domain-invariant feature learning, sample selection bias, and hyperparameter optimization.

When applying a model trained on one domain (source) to another (target), a \textbf{\textit{distribution shift}} often occurs, referred to as domain shift. Formally, if \(P_S(X, Y)\) and \(P_T(X, Y)\) represent the joint distributions of features \(X\) and labels \(Y\) in the source and target domains, respectively, the challenge arises when \(P_S(X, Y) \neq P_T(X, Y)\). To address this, techniques focus on learning a transformation of the feature space to minimize the difference between the source and target distributions. This involves finding a mapping function \(f: X \rightarrow Z\), where \(Z\) is a latent space in which the distributions of transformed features \(f(X_S)\) and \(f(X_T)\) are more similar, quantified using measures such as the \ac{MMD}. \textbf{\textit{Label distribution shift occurs}} when the distributions of labels (\(P_S(Y)\) vs. \(P_T(Y)\)) differ, even if the feature distributions align. This poses challenges, especially with underrepresented classes in the target domain. Addressing this mathematically involves adjusting the model or learning process, possibly by re-weighting the loss function based on class distribution estimates. \textit{\textbf{Catastrophic forgetting}} is a risk during fine-tuning on a new domain, where the model may lose its performance on the original task. Balancing loss functions (\(L_S\) for the source and \(L_T\) for the target) is crucial, often weighted by a hyperparameter \(\lambda\) to control their importance. \textit{\textbf{Domain-invariant feature learning}} aims to learn features invariant across domains while remaining predictive. It involves optimizing a feature extractor \(f\) and predictor \(g\) to minimize the $D_S$ loss \(L_S\) and the domain discrepancy (e.g., \ac{MMD}). The problem of \textit{\textbf{sample selection bias}} occurs in selecting samples for \ac{DA}, affecting the effectiveness of adaptation strategies. Mathematically, addressing this bias involves weighting or selecting samples to minimize it, often using importance sampling or re-weighting techniques. \textit{\textbf{Hyperparameter optimization}} is critical in \ac{DA}, where the choice of hyperparameters (e.g., \(\lambda\)) significantly impacts performance. Finding the optimal hyperparameters typically involves solving complex optimization problems using techniques like grid search, random search, or Bayesian optimization on a validation set.

Moreover, \textit{\textbf{the unification of \ac{DTL}}} in \ac{ASR} poses a challenge due to the varied mathematical formulations used in different studies. While efforts
have been made to unify definitions and formulations, further work is needed for a consistent understanding of \ac{DTL}. Speech-based \ac{DTL} processing faces challenges compared to image-based processing due to potential mismatches between source and target databases arising from factors like language, speakers, age groups, ethnicity, and acoustic environments. The \ac{CTC} approach, while promising, is limited by the assumption of frame independence. Cross-lingual \ac{DTL} challenges include incorporating linguistic characteristics from multiple sources and integrating knowledge at different hierarchical levels, considering linguistic differences. Finally, \textbf{\textit{computational burden}} remains a significant challenge in \ac{DTL} and \ac{DA} processes. Knowledge transfer between domains can incur additional computational costs, especially considering the extensive computational resources required for deep architectures inherent in \ac{DTL} techniques.

\subsection{FL-based}
\ac{FL} has significant potential for \ac{DTL} systems, particularly in enhancing privacy and personalization. However, deploying this technology in \ac{DTL} also comes with a set of challenges. 
Typically, in \ac{FL}, data is inherently decentralized and can vary significantly across devices. This \textbf{\textit{heterogeneity in speech}} data—due to differences in accents, dialects, languages, and background noise—can make it challenging to train a model that performs well across all nodes. Ensuring robustness and generalization of the \ac{DTL} model under these conditions is a complex task. Moving on, \ac{FL} requires periodic communication between the central server and the devices to update the model. For \ac{DTL} systems, where models can be quite large, this can result in \textit{\textbf{substantial communication overhead}}. Optimizing the efficiency of these updates, in terms of both bandwidth usage and energy consumption, especially on mobile devices, is a significant challenge. Besides, although \ac{FL} is designed to enhance privacy by not sharing raw data, there are still \textit{\textbf{privacy challenges}}. For instance, it's possible to infer sensitive information from model updates. Ensuring that these updates do not leak private information about the users' speech data is a critical concern that requires sophisticated privacy-preserving techniques like differential privacy or secure multi-party computation.

Additionally, one of the advantages of \ac{FL} is the ability to personalize models based on local data. However, \textbf{\textit{balancing personalization}} with the need for a generally effective model—especially in a diverse ecosystem with varying speech patterns—is challenging. Achieving this balance without compromising the model's overall performance or the personalization benefits is a key challenge. Moving forward, \ac{FL} systems need to manage potentially thousands or millions of devices participating in the training process. \textit{\textbf{Scalability issues}}, including managing updates from such a large and potentially unreliable network of devices, ensuring consistent model improvements, and handling devices joining or leaving the network, are significant technical hurdles. Lastly, in \ac{FL}, the distribution of data across devices is often \ac{non-IID}. This means that the speech data on one device might be very different from that on another, leading to challenges in training a model that generalizes well across all devices. Overcoming the bias introduced by \textit{\textbf{\ac{non-IID} data}} is a major challenge in \ac{FL} for \ac{ASR}.

\subsection{DRL-based}
Using \ac{DRL} in \ac{ASR} systems offers promising avenues for improvement but also presents several challenges. Specifically, one of the primary challenges in applying \ac{DRL} to \ac{ASR} is the issue of \textit{\textbf{sparse and delayed rewards}}. In many \ac{ASR} tasks, the system only receives feedback (rewards or penalties) after processing lengthy sequences of speech, making it difficult to attribute the reward to specific actions or decisions. This delay complicates the learning process, as the model struggles to identify which actions led to successful outcomes.
Moreover, \textit{\textbf{balancing exploration}}, trying new actions to discover their effects, with exploitation, using known actions that yield the best results, is a critical challenge in \ac{DRL}. In the context of \ac{ASR}, this means the system must balance between adhering to known speech patterns and exploring new patterns or interpretations. Overemphasis on exploration can lead to inaccurate transcriptions, while excessive exploitation may prevent the model from adapting to new speakers or accents. Additionally, \ac{DRL} models typically require a significant amount of interaction data to learn effectively. In \ac{ASR}, obtaining \textit{\textbf{large volumes of labeled speech}} data, especially with user feedback, can be challenging and expensive. Additionally, \ac{DRL} algorithms can be sample-inefficient, meaning they need a lot of data before they start performing well, which can be a \textbf{\textit{bottleneck}} in practical applications.

Moving forward, most \ac{ASR} systems are built using supervised learning techniques that rely on vast amounts of annotated data. Integrating \ac{DRL} into these systems poses technical challenges, as it requires a different training paradigm that focuses on learning from user interactions and feedback rather than static datasets. Besides, using \ac{DRL} in \ac{ASR} often involves collecting and analyzing user feedback and interactions to improve the model. This raises concerns about \textit{\textbf{user privacy and data security}}, as sensitive information might be inadvertently captured and used for training. Ensuring that data is handled securely and in compliance with privacy regulations is a significant challenge.

Designing an \textit{\textbf{appropriate reward}} function that accurately reflects the desired outcomes in \ac{ASR} is challenging. The reward function must capture the nuances of speech recognition, such as accuracy, naturalness, and user satisfaction, which can be difficult to quantify. Poorly designed reward functions can lead to suboptimal learning outcomes or unintended behaviors. Lastly, \ac{ASR} systems are used in a wide range of environments, from quiet offices to noisy streets. \ac{DRL} models need to adapt to these \textit{\textbf{varying conditions}}, but training them to handle such diversity can be complex. The environment's variability requires models that can generalize well across different acoustic conditions, which remains a challenge for \ac{DRL}-based \ac{ASR} systems.

\section{Future directions} \label{sec5}

\subsection{Personalized data augmentation for dysarthric and older people}
While \ac{DTL} technologies have advanced significantly, especially in recognizing typical speech patterns, they still struggle to accurately identify speech from individuals with dysarthria or older adults \cite{hamza2023machine}. Gathering extensive datasets from these groups is challenging due to mobility limitations often associated with these populations. In this context, personalized data augmentation plays a crucial role \cite{feng2024towards,zhou2024adversarial}. 
Personalized data augmentation tailors the training process to accommodate the unique speech patterns and challenges associated with these groups. Dysarthria, a motor speech disorder, and the natural aging process can lead to speech that deviates from the normative models typically used to train \ac{DTL} systems, making accurate recognition difficult \cite{wei2024adastreamlite}. Personalized data augmentation introduces a wider range of speech variations into the training dataset, including those specific to dysarthric speakers or older adults. This can include variations in speech rate, pitch, articulation, and clarity. By training on this augmented dataset, the \ac{DTL} system learns to recognize and accurately transcribe speech that exhibits these characteristics \cite{yeo2024akvsr}.
Moreover, this helps the \ac{DTL} models generalize better to unseen examples of speech from dysarthric speakers or older adults. This enhanced generalization is crucial for real-world applications where the system encounters a wide range of speech variations. Moving forward, personalized data augmentation can employ specific techniques tailored to the needs of dysarthric speakers or older adults, such as simulating the slurring of words, varying speech tempo, or introducing background noise \cite {djeffal2023noise}, commonly challenging for these groups \cite{zhao2024driver}. Techniques like pitch perturbation, temporal stretching, and adding noise can simulate real-world conditions more accurately for these users.
A typical example is presented in \cite{jin2023personalized}, where a unique approach utilizes speaker-dependent \acp{GAN} has been proposed.

\subsection{Multitask learning for ASR}
\Ac{MTL} enhances the performance of \ac{DTL} systems by leveraging the inherent relatedness of multiple learning tasks to improve the generalization of the primary \ac{ASR} task. This approach allows the \ac{ASR} model to learn shared representations that capture underlying patterns across different but related tasks, leading to several key benefits \cite{brack2024sequential}. Typically,  \ac{MTL} encourages the \ac{ASR} system to learn representations that are beneficial across multiple tasks. This can lead to more robust feature extraction, as the model is not optimized solely for transcribing speech but also for other related tasks, such as speaker identification or emotion recognition. This shared learning process helps in capturing a broader range of speech characteristics, which can improve the \ac{ASR} system's ability to handle varied speech inputs. Moving on, by simultaneously learning related tasks, \ac{MTL} acts as a form of regularization, reducing the risk of overfitting on the primary \ac{ASR} task. This is because the model must find a solution that performs well across all tasks, preventing it from relying too heavily on noise or idiosyncrasies specific to the training data of the main task. Besides, learning auxiliary tasks alongside the main \ac{ASR} task can improve the model's generalization capabilities. For example, learning to identify the speaker or the language can provide additional contextual clues that help the \ac{ASR} system better understand and transcribe ambiguous audio signals.

Additionally, \ac{MTL} can make more efficient use of available data by leveraging auxiliary tasks for which more data might be available. In scenarios where annotated data for \ac{ASR} is limited, incorporating additional tasks with more abundant data can help improve the learning process and performance of the \ac{ASR} system. Moreover, \ac{MTL} allows \ac{ASR} systems to better handle acoustic variability in speech, such as accents, dialects, or noisy environments, by incorporating tasks that directly or indirectly encourage the model to learn features that are invariant to these variations. Last but not least, modern \ac{ASR} systems often employ \ac{DL} architectures that can benefit from end-to-end learning strategies. \ac{MTL} fits naturally into this paradigm, allowing for the joint optimization of multiple objectives within a single model architecture. This can simplify the training process and reduce the need for separately trained models or handcrafted features.

\subsection{Federated multi-task learning and distillation for ASR}
\Ac{FMTL} extends the concept of \ac{FL} by allowing each client to learn a personalized model that addresses its specific task, while still benefiting from collaboration with other clients. This approach recognizes the heterogeneity in clients' data distributions and tasks. Mathematically and compared with \ac{FL} (Equation \ref{FL}), \ac{FMTL} can be formulated as:

\begin{equation}
  \min_{\theta_1, \theta_2, ..., \theta_K} \sum_{k=1}^{K} F_k(\theta_k) + \lambda R(\theta_1, \theta_2, ..., \theta_K)  
\end{equation}

Different from \ac{FL}, $ R(\theta_1, \theta_2, ..., \theta_K) $ is a regularization term that encourages some form of similarity or sharing among the model parameters of different tasks, promoting collaboration among clients. $ \lambda $ is a regularization coefficient that balances the trade-off between fitting the local data well and collaborating with other clients. \ac{FMTL} has task-specific model parameters $ \theta_k $ for each client, where only a single global model parameter $ \theta $ in \ac{FL}.

In this regard, \ac{FMTL} offers a promising approach to improving \ac{ASR} systems while also enhancing privacy and security measures. This learning paradigm extends the traditional \ac{FL} model by enabling the simultaneous training of multiple tasks across distributed devices or nodes, without the need to share raw data \cite{zhang2023federated}. \ac{FMTL} leverages data from a wide range of devices and users, each potentially offering unique speech data, accents, dialects, and noise conditions. This diversity helps in training more robust \ac{ASR} models that can perform well across various speech patterns and environments \cite{singh2023federated}. By learning from many tasks simultaneously, \ac{FMTL} can personalize \ac{ASR} models to individual users or specific groups without compromising the model's general performance \cite{jiang2023fedradar}. This is particularly beneficial for users with unique speech patterns, such as those with accents or speech impairments. Moreover, \ac{FMTL} encourages the development of compact models that can handle multiple tasks efficiently. For \ac{ASR} systems, this means that a single model can potentially perform speech recognition, speaker identification, and even emotion detection, reducing the computational overhead on client devices \cite{azadi2024robust}.

On the other hand, in \ac{FMTL}, raw data remains on the user's device and does not need to be shared or transferred to a central server. This inherently reduces the risk of data breaches and unauthorized access, as sensitive speech data is not centralized \cite{ji2024edge}. Additionally, \ac{FMTL} can be combined with differential privacy techniques to further anonymize the model updates sent from devices to the central server. This ensures that the shared information does not reveal sensitive details about the data or the user, enhancing privacy protection \cite{vsajina2024multi}. Moving on, the aggregation process in \ac{FMTL} can be secured using cryptographic techniques, ensuring that the aggregated model updates cannot be traced back to individual users. This secure aggregation process protects user privacy while allowing the benefits of collective learning \cite{ye2023pfedsa}. Lastly, by aggregating model updates from a wide range of tasks and users, \ac{FMTL} can improve the system's robustness to malicious attempts at data poisoning. The diversity of inputs helps in diluting the impact of any adversarial data introduced to compromise the model.

Delving deeper into techniques for \ac{FL} distillation (optimizing model size) within \ac{FL} frameworks is essential. This exploration involves researching methods to compress neural \ac{ASR} models effectively while ensuring their performance remains intact, especially tailored for edge devices with storage and computational constraints. It is imperative to investigate the trade-offs associated with reducing model size while maintaining performance metrics like \ac{WER}. Developing strategies to strike a balance between downsizing models and preserving satisfactory performance levels within \ac{FL} environments is crucial.

\subsection{Recent  DRL techniques for \ac{ASR}}
Exploring  incremental \ac{DRL} approaches \cite{wang2021lifelong, wang2020irda, wang2019incremental} in \ac{DRL}-based \ac{ASR} systems,  could be very interesting. This approach involves the model continuously learning from newly acquired data and dynamically adjusting its ASR functionalities over time. By incrementally updating its knowledge base, the model can enhance its performance without necessitating full retraining, thus enabling continual enhancement of \ac{ASR} systems. This capability not only fosters greater resilience and adaptability in speech recognition capabilities but also offers potential applications in scenarios where real-time adaptation to changing conditions is crucial, such as in noisy environments or with varying speaker accents. Moreover, incremental \ac{DRL} can potentially lead to more efficient use of computational resources, as the model only needs to focus on learning from new data, rather than reprocessing the entire dataset. Further research in this area could unlock new possibilities for \ac{ASR} systems to evolve and improve over time, ultimately enhancing their usability and effectiveness in diverse real-world settings.

Although some \ac{DTL} schemes based on \ac{DRL} have been proposed, there remains a notable scarcity in the application of \ac{DRL} techniques to enhance \ac{DTL} methods. While policy gradient and Q-learning are commonly employed, the realm of \ac{DRL} encompasses various subcategories such as  \ac{DDQN}, \ac{AC}, \ac{DDPG}, and more \cite{gueriani2023deep}, which hold promise for advancing \ac{DTL} with innovative approaches. Researchers are encouraged to delve into these diverse \ac{DRL}-based methods to further enrich the field of \ac{DTL} for both \ac{AM} and \ac{LM} fields.

\subsection{Online \ac{DTL}}
Online \ac{DTL} combines the principles of DTL and online learning with \acp{DNN}, enabling models to adapt in real-time to new tasks or data distributions. This approach is beneficial in dynamic environments where data arrives sequentially. \ac{DL} models, specifically \acp{DNN}, learn through optimizing the weights $\theta$ to minimize a loss function $L$, which measures the discrepancy between predicted outputs $\hat{y}$ and true outputs $y$:  $\theta^* = \arg\min_\theta L(D; \theta)$. On the other hand, DTL improves learning in a new target task through the transfer of knowledge from a related source task, adapting a pre-trained model $\theta_S$ on $D_S$ to a $D_T$: $ \theta_T^* = \arg\min_\theta L(D_T; \theta_T)$. In this regard, online learning updates the model incrementally as new data $(x_t, y_t)$ arrives:

\begin{equation}
    \theta_{t+1} = \theta_t - \alpha_t \nabla_\theta L(y_t, f(x_t; \theta_t))
\end{equation}

\noindent Where $\alpha_t$ is the learning rate, and $\nabla_\theta L$ is the gradient of the loss with respect to $\theta$. Moving on, online \ac{DTL} integrates these concepts to continuously adapt a deep learning model to new tasks or data streams, often involving techniques such as feature extraction, fine-tuning, model adaptation, and continual learning. The adaptation process at each time step $t$ can be viewed as $\theta_{t+1}^* = \arg\min_\theta L(D_{T_t}; \theta_{T_t})$.  Where $D_{T_t}$ represents the data available at time $t$, including new target domain data.

The approach of online \ac{DTL} offers a forward-looking solution to this issue. Particularly, discrepancies in class distributions and the representation of features between the $D_S$ and $D_T$ amplify the complexity of online \ac{DTL} \cite{zhao2014online}.
To navigate the complexities mentioned, online \ac{DTL} has been dissected into two primary methodologies. The first, known as homogeneous online \ac{DTL}, operates on the premise of a unified feature space across both domains. Conversely, heterogeneous online \ac{DTL} acknowledges the distinct feature spaces intrinsic to each domain \cite{wu2017online}. An exemplary solution to the challenges of heterogeneous online \ac{DTL} includes leveraging unlabeled instances of co-occurrence to forge a connective bridge between the $D_S$ and $D_T$, facilitating the precursor to knowledge transfer \cite{wu2019online}.
Furthering the discourse, online \ac{DTL} augmented with extreme learning machines introduces a novel framework \cite{alasbahi2022an}. Addressing the challenge of limited data in the $D_T$, the technique of \ac{DTL} with lag, rooted in shallow neural network embeddings, has been applied. This method ensures the continuity of knowledge transfer, notwithstanding fluctuations in the feature set.

\subsection{Transformers and LLMs-based ASR}

\Acp{LLM} and Transformers represent the forefront of \ac{AI}, trained on vast datasets spanning various domains, including text, speech, images, and multi-modal inputs. Despite extensive research on \ac{ASR}, existing \ac{SOTA} approaches often lack integration of advanced \ac{AI} techniques like \ac{DRL} and \ac{FL} into both \ac{AM} and \ac{LM} domains.   For example, \Ac{LLM} based on \ac{DTL}  has demonstrated significant potential for \ac{ASR} tasks, particularly for both \ac{LM} and \ac{AM} components. The incorporation of \ac{DTL} techniques into \ac{LLM} facilitates the transfer of knowledge from extensive pre-training tasks to enhance \ac{ASR} effectiveness. In terms of \ac{AM}, fine-tuning \ac{LLM}  can leverage insights gained from pre-trained models exposed to sample acoustic data. This enables the \ac{AM} component to grasp acoustic features like spectrograms or Mel-frequency cepstral coefficients (MFCCs)  and utilize pre-trained knowledge to improve speech recognition accuracy. Through fine-tuning, the model can adjust and specialize in specific datasets or acoustic domains, leading to enhanced \ac{ASR} performance. 

Similarly, the \ac{LM} aspect of \ac{LLM} based on \ac{DTL} can enhance \ac{ASR} by leveraging \ac{DTL}. Pre-training \ac{LLM} on vast text corpora equips it with extensive language representations, aiding in addressing diverse language challenges. Fine-tuning enables adaptation to specific language characteristics, improving transcription accuracy and contextual appropriateness. Both \ac{AM} and \ac{LM} fine-tuning can benefit from \ac{DA}, incorporating target domain data to tailor models, reducing domain mismatch, and enhancing effectiveness and generalization. Utilizing \ac{LLM} for objective \ac{ASR} testing and MOS evaluation involves compiling diverse datasets, fine-tuning \ac{LLM}, and integrating it into the \ac{ASR} system. Evaluation metrics like \ac{CER} and Pearson's correlation gauge system performance, guiding further fine-tuning iterations for improved results. This iterative process ensures the \ac{ASR} system's continual enhancement and accurate MOS scale generation. Researchers are invited to explore these gaps and advance the integration of more advanced \ac{AI} techniques such as \ac{DRL} and \ac{FL} into both \ac{AM} and \ac{LM} domains within Transformer-based models. Additionally, there is a need for further investigation into Transformers and \ac{DTL}-based ASR schemes specifically tailored for the \ac{LM} domain. Closing these gaps will contribute to the development of more robust and effective language models across various applications.

\section{Conclusion} \label{sec6}

In conclusion, recent advancements in deep learning have presented both challenges and opportunities for \ac{ASR}. Traditional \ac{ASR} systems require extensive training datasets, often including confidential information, and consume significant computational resources. However, the demand for adaptive systems capable of performing well in dynamic environments has spurred the development of advanced deep learning techniques such as \ac{DTL}, \ac{FL}, and \ac{DRL}, with all their variant techniques. These advanced techniques address issues related to \ac{DA}, privacy preservation, and dynamic decision-making, thereby enhancing \ac{ASR} performance and reducing computational costs.

This survey has provided a comprehensive review of \ac{DTL}, \ac{FL}, and \ac{DRL}-based \ac{DTL} frameworks, offering insights into the latest developments and helping researchers and professionals understand current challenges. Additionally, the integration of Transformers, powerful \ac{DL} models, has been explored for their ability to capture complex dependencies in \ac{DTL} sequences. By presenting a structured taxonomy and conducting critical analyses, this paper has shed light on the strengths and weaknesses of existing frameworks, as well as highlighted ongoing challenges. Moving forward, further research is needed to overcome these challenges and unlock the full potential of advanced DL techniques in \ac{DTL}. Future work should focus on refining existing approaches, addressing privacy concerns in \ac{FL}, improving \ac{DRL} algorithms for \ac{DTL} optimization, and exploring innovative ways to leverage Transformers for more efficient and accurate speech recognition. By continuing to innovate and collaborate across disciplines, we can push the boundaries of \ac{ASR} technology and realize its transformative impact on various fields, including healthcare, communication, and accessibility.

\balance
\bibliographystyle{elsarticle-num}
\bibliography{references}

\end{document}